Elsevier Editorial System(tm) for Physica A
Manuscript Draft

Manuscript Number: PHYSA-19818R1

Title: Complex Network Construction of Internet Financial Systemic Risk

Article Type: Research Paper

Section/Category: Networks

Corresponding Author: Professor Chuanmin Mi,

Corresponding Author's Institution:

First Author: Runjie Xu

Order of Authors: Runjie Xu; Chuanmin Mi; Rafał Mierzwiak; Runyu Meng

Abstract: Internet Finance is a new financial model that applies Internet technology to payment, capital borrowing and lending and transaction processing. In order to study the internal risks, this paper uses the Internet Finance risk elements as the network node to construct the complex network of Internet Finance risk system. Different from the study of macroeconomic shocks and Finance institution data, this paper mainly adopts the perspective of complex system to analyze the systematic risk of Internet Finance. By dividing the entire financial system into Internet Finance subnet, regulatory subnet and traditional financial subnet, the paper discusses the relationship between contagion and contagion among different risk factors, and concludes that risks are transmitted externally through the internal circulation of Internet Finance, thus discovering potential hidden dangers of systemic risks. The results show that the nodes around the center of the whole system are the main objects of financial risk contagion in the Internet Finance network. In addition, macro-prudential regulation plays a decisive role in the control of the Internet Finance system, and points out the reasons why the current regulatory measures are still limited. This paper summarizes a research model which is still in its infancy, hoping to open up new prospects and directions for us to understand the cascading behaviors of Internet Finance risks.

# Complex Network Construction of Internet Finance Risk

Runjie Xu[a], Chuanmin Mi[a*] , Rafał Mierzwiak[b], RunYu Meng[c]

[a] College of Economics and Management, Nanjing University of Aeronautics and Astronautics

Nanjing, Jiangsu 210016, China

[b] Faculty of Engineering Management, Poznan University of Technology, 60-965 Poznan, Poland

[c] University of Science and Technology of China, Anhui, Hefei, 230026, China

**Highlights**

- Construct a complex network with the risk elements of Internet Finance, and propose a systemic risk contagion system based on Internet Finance, traditional Finance and supervision.
- Propose the comprehensive importance, contagion intensity and vulnerability sensitivity of Internet Finance risk factors, and describes the contagion cascade behavior in the system.
- Use topological images to predict the possibility of Internet Finance inducing systemic risk, and to judge other potential risks of systemic risk.

**Abstract**

Internet Finance is a new financial model that applies Internet technology to payment, capital borrowing and lending and transaction processing. In order to study the internal risks, this paper uses the Internet Finance risk elements as the network node to construct the complex network of Internet Finance risk system. Different from the study of macroeconomic shocks and Finance institution data, this paper mainly adopts the perspective of complex system to analyze the systematic risk of Internet Finance. By dividing the entire financial system into Internet Finance subnet, regulatory subnet and traditional financial subnet, the paper discusses the relationship between contagion and contagion among different risk factors, and concludes that risks are transmitted externally through the internal circulation of Internet Finance, thus discovering potential hidden dangers of systemic risks. The results show that the nodes around the center of the whole system are the main objects of financial risk contagion in the Internet Finance network. In addition, macro-prudential regulation plays a decisive role in the control of the Internet Finance system, and points out the reasons why the current regulatory measures are still limited. This paper summarizes a research model which is still in its infancy, hoping to open up new prospects and directions for us to understand the cascading behaviors of Internet Finance risks.

## 1.Introduction

Internet Finance is an organic combination of financial industry and Internet information technology. Internet Finance, represented by mobile payment, online loan, crowdfunding and Internet investment, has developed vigorously and become a hot spot of financial innovation. The development of Internet Finance not only improves the efficiency of financial resource allocation and realizes the transformation of residents' savings into capital, but also has certain risk attributes, which has attracted the attention of regulatory authorities. For example, the innovation vitality of Internet Finance products forces the current relevant laws and regulations still lag behind; The failure of Internet Finance platform causes frequent capital problems; Some large-scale Internet Finance platforms have gradually become the core of the Internet Finance system and have the majority of users. The impact of Internet Finance on the economy has an important impact on the robustness of the entire financial system and the ability to serve the real economy. Therefore, the development of Internet Finance puts forward higher requirements for financial supervision departments. How to further prevent and resolve Internet Finance risks while promoting the development of Internet Finance has become an important issue to be solved.

### 1.1 Related work

According to the existing research, the study of financial systemic risk is usually to analyze the impact of financial institution size, lending behavior, risk ability and other factors on risk contagion. There are a series of models proposed in the literature, such as the conditional risk value model (CoVaR) [1], the crisis insurance cost model (DIP)[2], the systematic expected shortage model (SES) [3], and the systematic or conditional analysis method (SCCA ) [4], etc. However, Internet Finance, as a new financial mode, has to take into account the risk of contagion and spontaneity, which is more prominent than the traditional financial industry. For example, Lee and Lee empirically studied the existence of herding behavior in network P2P loans, which indicates that Internet Finance is also full of media attributes, which makes the reputation of Internet Finance industry to have the ability of fast influence of herding effect [5]. Suresh Kotha et al. studied three types of reputation building activities using the top 50 pure Internet companies as a sample [6]. Reputation-building activities may be one of the key determinants of the success of Internet competition.

However, the lack of data limits the study of financial systemic risk to some countries [7, 8]. Moreover, in the face of such a complex system of financial risks, the perspective of financial institutions' business transactions or macroeconomic pressures has been stretched. Therefore, the study of systemic risk requires a more extensive approach, that is, the use of interactions in the financial system to study systemic risk [9-11]. Some scholars have showed that the risk transfer and the evolution mechanism in the system by constructing the network form [12]. The network topology is used to solve the problems in the financial network. Systemic risk drivers, applied to cross-country exposure data in BIS databases, Solorzano-Margain et al. use network theory to describe the contagion of financial crisis [13]. Amini and Minca proposed a framework to test the risk of large-scale cascading crisis in financial network and studied the contagion problem in financial network [14]. Alex et al. concluded that risk mitigation and optimal repair were largely dependent on the interdependent network structure, and the problem of monitoring risk was solved through the network dependence perspective[15].

### 1.2 Method

In many types of network construction methods, the cross penetration and close combination of complex network theory between different disciplines have proved its excellent ability in analyzing and processing complex systems. The technology of network science has been successfully applied to the analysis of financial system [16-18]. Based on this, this paper uses complex networks to describe the geometric properties of Internet Finance risk system, analyzes the formation mechanism of the system and predicts the structural stability of the system, and then explains the macro and micro characteristics of an Internet Finance system itself. This approach brought great practical significance for the study of the spread of systemic risks in Internet Finance. Through the theory of complex networks, this paper makes use of the different inductive factors of Internet Finance risk and its influence relationship, and builds the network step by step, thus exploring the contagion path and network characteristics under Internet Finance risk. The results show that Internet Finance has the potential to induce systemic risk, and its important source lies in the external effect caused by the significant risks imposed on the outside world by the core Internet Finance institutions. In addition, the diffusion mechanism of Internet Finance risk is different from the traditional financial network in the past. The development of Internet Finance not only improves the efficiency of financial resource allocation and realizes the transformation of residents' savings into capital, but also has certain risk attributes, which has attracted the attention of regulatory authorities. Stronger core influence and ability to change, which also puts forward more targeted requirements for regulators. In this paper, the research on Internet Finance based on the theory of complexity science can become a new idea, new method and new way to realize the unity of micro and macro research on Internet Finance. For the modern Internet Finance regulation and detection of potential dangers to provide more effective support. The method could also provide inspirations for other scientific areas, such as death danger from organ failure in Biology, system breakdown risks in Ecology, overload hazard in computational network.

## 2. Risk Relationships Build Networks

At present, there are two main methods to construct financial risk contagion network: one is to establish the network according to the actual lending data or the actual transaction data of payment system between financial institutions [19-22], the other is based on the stock market information such as stock price, daily rate of return to establish a network [23-25]. According to the statistical characteristics of network geometry, the complex network models mainly include regular network model, random network model, W-S small-world network model [26] and B-A scale-free network model [27]. Newman et al. think that complex network is a higher-level relational network composed of multi-level networks [28]. Anything in real life can be understood as a complex network of different levels. Strogatz shows that a complex network is composed of interrelated nodes to describe a variety of real complex systems [29]. Based on this, this paper consider the Internet Finance Risk as a complex system. It uses different risk factors as nodes to construct a network model, and uses the number of connected nodes of risk factors to predict risk impact ability.

Assume the network feature to be composed of two basic elements: node $v$ and correlation mode $e$. If $V$ is regarded as the set of $v$, $E$ is regarded as the set of $e$, and each edge $e$ in $E$ has a pair of points $(i,j)$ corresponding to $V$. Then the whole network can be represented by the matrix $G = G(V,E)$, with elements defined by:

$$g_{ij} = \begin{cases} 1, \text{risk "i" will lead to the direct occurrence of risk "j", } i \geq 1 \\ 0, \text{otherwise} \end{cases} . \tag{1}$$

At the same time, because the complex system is a typical directed network, the matrix can be judged as an asymmetric matrix, which means:

$$g_{ij} = g_{ji}. \tag{2}$$

For the degree "$k_i$" of node "$v_i$" in the network, it is defined as the number of connected edges:

$$k_i = \sum_{j=1}^{n} g_{ij}. \tag{3}$$

The overall importance of a node. Considering the entire Internet Finance risk system, risk factors will not be fixed, but will change with the development of the Internet. Therefore, it is assumed that at the initial time $t = 0$, the system has $m_0$ risk factors, and at each subsequent time interval, node $(m \leq m_0)$ with degree m will keep changing. The probability of a node with degree m connecting to other nodes is: $P(k_i)$, which is proportional to the degree $k_i$ of the original node. In the real Internet Finance system, it is explained as follows: the factor with stronger risk contagion ability has stronger contagion ability to the outside world, and the risk factor with weaker risk resistance ability is more likely to be infected.

The probability that a node has a connection edge is: $P(k_i) = \dfrac{k_i}{\sum_j k_j}$.

This value is defined as the cumulative risk that the node can accumulate when the whole network is attacked, and the ability to infect the risk to other nodes, which is a comprehensive evaluation of different Internet Finance risk elements.

Similarly, the probability that a node has a continuous edge is: $P(k_i)^{\rightarrow} = \dfrac{k_i^{\rightarrow}}{\sum_j k_j}$.

This value represents the infection intensity of a node. From the analysis of the evolution mechanism of Internet Finance risks, it is assumed that there is only one risk source I before the unstable state of the Internet Finance system. If the risk source $i$ will lead to the subsequent occurrence of other risks, the stronger the influence of risk source $i$ is, that is, the higher the $P(k_i)^{\rightarrow}$ value is. This indicates that the more likely a risk source is to play the role of inducing systemic risk in the system, the more attention it needs to be paid. On the contrary, a low value indicates that a node has a low influence on the outside world.

However, the true contagion of risk sources must also take into account the vulnerability of other nodes in the Internet Finance system. Therefore, there is also a need for the probability of nodes being connected: $P(k_i)^{\leftarrow} = \dfrac{k_i^{\leftarrow}}{\sum_j k_j}$.

This value represents the sensitivity of a node to attack. The larger the value is, the more vulnerable the node is to other risks, and the less vulnerable to external infection.

It is found that the financial risk of Internet has local aggregation in the process of contagion, which is shown by the fact that part of the network is closely connected, and part of the connection is sparse, that is, the network can be divided into multiple subnetworks. This phenomenon exists in many real networks, such as the conclusion of "community structure" proposed by Newman [30, 31] that the systemic risk of finance has the nature of multi-layer network. Therefore, this paper divides the systemic risk of Internet Finance into three layers: Internet

Finance network, regulatory network and traditional financial network. Integrating the characteristics of Internet Finance and traditional financial industry, this paper starts with the three elements of "causes of risk induction", "means of risk regulation" and "path of risk contagion" in the process of network construction. First of all, we pay attention to the source of systemic risk induced by Internet Finance, and analyze the risk factors in the whole Internet Finance, and study the cause of risk contagion and the path of contagion. Secondly, considering the business activities between regulatory authorities and Internet Finance, capital flow and risk exposure, using the flow relationship between the two, regulatory intensity, regulatory costs and other constraints to build a complex system of regulatory networks. Finally, the paper considers the risks of traditional finance to construct traditional finance Subnet. Therefore, the network of Internet Finance risk system with multi-source risk superposition and multi-risk level can be represented synthetically.

The core mechanism of network construction in this paper is to judge whether there is a direct inductive relationship between different risk factors, and if there is direct infection, the directed routes of both infectious parties will be established. For the judgment of inducement and trigger between risk factors, this paper mainly through the history of the Internet industry and the existing literature theory synthesis. The conduction relationship between factors is shown in Fig. 1 below.

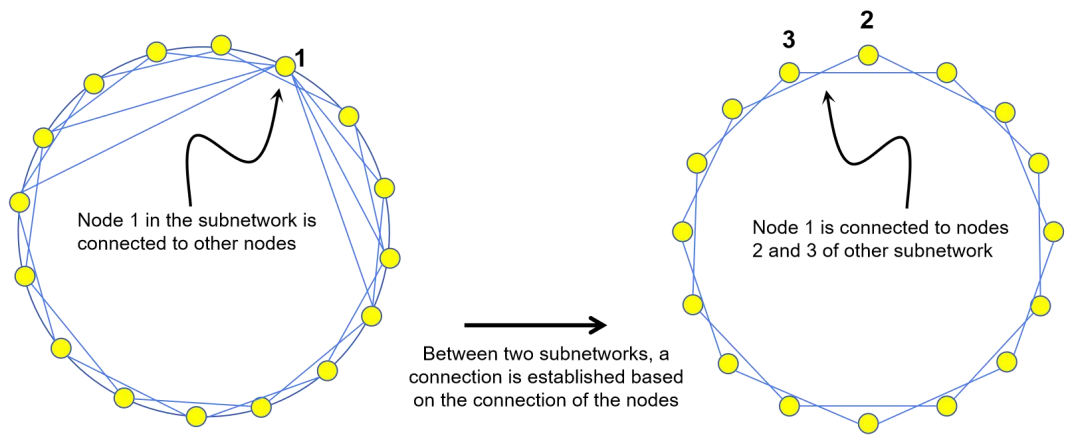

**Fig. 1.** Represents the structural relationship between two risk network layers. In the graph, node 1 has a risk contagion in its subnetwork, and node 1 is also associated with node 2 and node 3 in another subnetwork, so the association between different subnets is based on the interconnection between nodes. In reality, the internal transaction disorder of an Internet Finance platform will not only cause the crisis of the Internet Finance platform itself, but also affect the final settlement bank.

After the analysis of the network structure, the graph layout algorithm can show the scattered information in a clear way, and meet the corresponding aesthetic standards. Therefore, this paper adopts two force guided placement algorithms, Force Atlas and Fruchterman Reingold, which mimic the gravity and repulsion of the physical world, automatically layout until the force is balanced, and make the graph more compact and more readable. After the complex network connection is transformed into a more beautiful network layout, it is convenient to observe the overall structure of the network and its automorphism characteristics.

## 3. Internet Finance Subnet

It should be emphasized that it is a complicated work to build a network for the entire Internet Finance risk system. The Internet Finance system not only involves its own risks, but also has a close relationship with commercial banks and regulators. We first determined the hierarchical network construction idea of the entire Internet Finance system, and then integrated a large number of literature, and through the investigation of actual bank workers to determine whether there is a correlation between risk factors and risk factors.

Comprehensive financial risk and Internet Finance risk research literature is provided by a series of papers [32-37]. Internet Finance risk mainly includes technical risk, operational risk and legal risk, credit risk and business risk. In this paper, the main types of risk in Internet Finance are decomposed into several important subdivision risks to construct the network of risk factors.

### 3.1 Technical risk

Technical risk include: operating system vulnerability factor, virus Trojan factor, internal information disclosure factor, identity forgery login factor, network transmission factor, server maintenance factor and natural disaster damage factor. Like the risk of traditional finance, each subdivision risk factor in Internet Finance is transmitted to each other [14]. For example, the operating system vulnerability will lead to Trojan horse virus intrusion, resulting in server malfunction and internal information disclosure, network transmission problems will lead to virus Trojan horse attacks and internal information disclosure. Improper server maintenance will cause network transmission problems and so on [38]. From the perspective of external effects, the type of technical risk will have cross-class contagion with other types of risk, mainly reflected in the impact on the type of operational risk and the type of legal risk. For example, operating system vulnerabilities, network transport problems, etc. can lead to malicious intrusion (operational risk type). Internal information disclosure may lead to the misuse of personal information (legal risk type) and user prosecution (legal risk type).

### 3.2 Operational risk

Operational risk mainly exists in the business model of Internet Finance, including internal operational risk factor, malicious intrusion risk factor, user accidental operation risk factor, service provider operating risk factor, outsourcing technology risk factors and cooperative development risk factors [35]. Part of the subdivision risk factors will be transmitted to each other. For example, service providers, outsourcing technology, or the process of cooperative development may lead to the deterioration of relationships, which may lead to the risk of malicious intrusions by partners. From the external impact, the type of operational risk will be cross-class contagion with other risk types, mainly reflected in the type of technical risk, legal risk type, industry. Service risk type, enterprise operation risk type. For example, after disputes that occur between Internet Finance enterprises, service providers, outsourcing partners, due to the limitation of the completeness of existing laws, it is not possible to effectively and timely investigate responsibilities, which may force the business development, reduce users' trust in it, and cause pressure on the operation of the whole enterprise.

### 3.3 Legal risk

Legal risk includes incomplete information disclosure risk factor, abuse of personal information risk factor, illegal financing risk factor, legal protection completeness risk factor, illegal business risk factor, users sue risk factors and national policy risk factors. Subdivision risk factors can be mutually transmitted. For example, the perfection of laws and regulations makes it impossible for Internet Finance companies to cover up their misuse of personal information, and they may also be subject to public relations crisis and user prosecution after the incident is exposed. From the external impact, the type of legal risk and other types of risk will be cross-class contagion, mainly reflected in the type of credit risk, business Risk type, enterprise operation risk type. For example, illegal Internet Finance platforms that illegally finance may have their own business model known as "Ponzi schemes". Compared with traditional finance, enterprises operate in a way that lacks industry norms and resists capital flows. Market cycle and interest rate risk ability is very low, platform easy credit breach risk.

**3.4 Credit risk**

Credit risk is ubiquitous in Internet Finance and is an important part of systematic risk prevention [19, 40]. This type of risk includes term mismatch risk factor, default contract risk factor, false publicity risk factor, etc. Platform running risk factor. Subdivision risk factors can be mutually transmitted. For example, Internet Finance products have a long term of investment assets and a short period of liabilities, which makes the financial products of Internet Finance enterprises unable to pay in time, thus creating the risk of term mismatch, which is the most general evolution into credit default. From the external impact, credit risk type and other types of risk will be cross-class contagion, the main Reflected in the legal risk type, business risk type, enterprise operation risk type. For example, after the risk of breach of contract on Internet Finance platform, there may also be prosecution or punishment by regulators, which reduces the confidence of Internet Finance business in the user's heart and thus affects the development of business activities [6].

**3.5 Business risk**

Business risk includes capital flow risk factor, market cycle risk factor, interest rate risk factor, user preference risk factor and investor relationship risk factor. Subdivision risk factors will have a certain degree of mutual transmission. For example, the inability of Internet Finance enterprises to obtain enough funds at reasonable cost and in a timely manner to cope with asset growth or the payment of maturing debts, resulting in the risk of a chain break of funds, is likely to be caused by investor relationships, it is also possible that interest rate risk, market cycle brought about. From the external impact, the type of business risk will be cross-class contagion with other types of risk, mainly reflected in the type of credit risk. Enterprise operating risk type. The performance of Internet Finance, for example, affects the health of its capital flows because many platforms attract users in the form of finance-subsidy in the process of doing business. This makes the platform inside the flow of capital health is extremely important.

**3.6 Management and strategy risk**

In addition, this paper also considers the healthy management and strategic choice of internet enterprises. No Internet Finance enterprise can avoid its own risks in operation management, expansion and strategic choice. Acquaah studied the importance of the business experience of business managers and the ability of community leaders in the use of resources [41]. Laeven & Levine made an empirical evaluation on the bank risk bearing theory, equity structure and national bank supervision [42]. It shows that the influence of supervision on bank risk depends on the corporate governance structure. In the process of regulating Internet Finance platforms, the marginal effect of risk is real and will change with the change of equity concentration. If the ownership structure of an Internet Finance platform is very unreasonable and CEO has great influence, the platform is very vulnerable to the influence of the personal will of CEO in the process of carrying out its business.

The healthy operation management and strategic choice of Internet enterprises mainly include the innovation vitality of enterprises, the competition mode in the industry, the public image of enterprises, the relationship of employees, the supplement of talents and posts, the salary level of employees, the decision-making of leadership, equity allocation and group parent-subsidiary relationship. These factors often cannot be quantified, and many Internet enterprises are difficult to collect relevant data, so it is rarely considered by the research of macro Internet Finance risk. By synthesizing the development of the Internet industry and existing research, this paper determines whether there is direct inducement between the risk factors, so that this kind of risk can be included in the financial network layer of the Internet analysis. According to the relationship between the subdivision risk, the subdivision risk factors will infect each other. For example, the leadership decision will affect the corporate image, talent introduction and salary level, and the salary level will affect the degree of talent introduction, and ultimately affect the innovation power of the Internet Finance platform. From the external impact, the types of operational risk mainly focus on the impact of leadership decisions. For example, whether the leadership and management will have the risk of internal operations, whether there will be illegal business and illegal financing decisions. The number and attributes of the subdividing risk under each type are shown in Table 1. In the Internet Finance subnet, force atlas model is used to visually arrange the entire network according to the sizes of $P(k_i)$, $\overleftarrow{P(k_i)}$ and $\overrightarrow{P(k_i)}$ of different risk nodes, as shown in Fig. 2 (a), (b) and (c).

**Table 1**

Six types of risks of Internet Finance are labeled as A (Technical risk type), B (Operation risk type), C (Legal risk type), D (Credit risk type), E (Business risk type). F (Operations and Strategic Options). Six types of risk were subdivided into a number of independent risks and labeled as A1, A2, A3 ······, F9

| Node name | Number | Node name | Number | Node name | Number |
|---|---|---|---|---|---|
| *Category A: Technical risks* | | | | | |
| System Vulnerability | A1 | Virus Trojan Horse | A2 | Internal Information Leakage | A3 |
| Identity Forgery Landing | A4 | Network Transmission | A5 | Server Maintenance | A6 |
| Natural Disaster Damage | A7 | | | | |
| *Category B: operational risk* | | | | | |
| Built-in Function | B1 | Malicious Intrusion | B2 | User Unexpected Operation | B3 |
| Service Provider Operation | B4 | Outsourcing Technology | B5 | Cooperate | B6 |
| *Category C: Legal risks* | | | | | |
| Information Disclosure | C1 | Individual Privacy | C2 | Illegal Financing | C3 |
| Incomplete Legal Protection | C4 | Illegal Operation | C5 | User Prosecution | C6 |
| National Policy | C7 | | | | |
| *Category D: Credit risk* | | | | | |
| Term Mismatch | D1 | Breach of Contract | D2 | False Propaganda | D3 |
| Platform Running | D4 | | | | |
| *Category E: Business risks* | | | | | |
| Capital Flow | E1 | Market Cycle | E2 | Interest Rate Exposure | E3 |
| User Preference | E4 | Investor Relationship | E5 | | |
| *Category F: Operation Management and Strategic Choice* | | | | | |
| Innovation Power | F1 | Competition Mode | F2 | Corporate Image | F3 |
| Employee Relationship | F4 | Talent Introduction | F5 | Salary Level | F6 |

| Leadership Decision-Making | F7 | Equity Allocation | F8 | Parent Subsidiary Company | F9 |

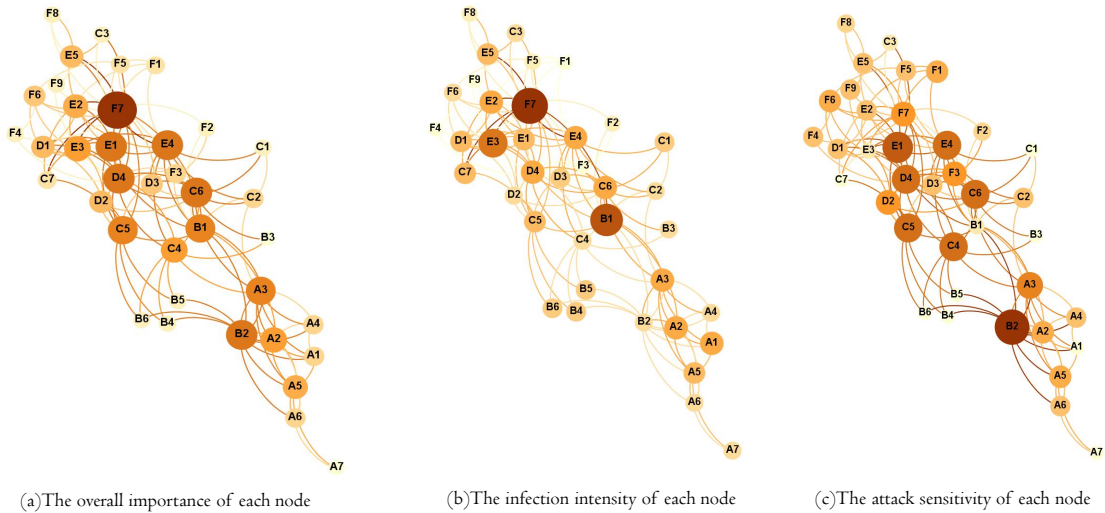

(a)The overall importance of each node    (b)The infection intensity of each node    (c)The attack sensitivity of each node

**Fig. 2** Comparison of the importance, contagion intensity, and sensitivity of each node in the Internet Finance subnet

The $P(k_i)$ of each node in Fig.2(a) is used as a measure of the comprehensive influence ability of nodes, which aims to highlight the factors that occupy important influence in the financial risk system of Internet. The images show that F7 nodes (Leadership Decision factors) are the most influential in the financial risk layer of the Internet, and the most influential of the six types of risk are the F node (Healthy Operation and Strategic Choice). The second is E node (Business risk category), C node (Legal risk category), D node (Credit risk category), B node (Operation risk category), A node (Technical risk category).

The $\overrightarrow{P(k_i)}$ shows the size of the link that each node emits, the intensity of contagion to the outside world. The six risk classes play a pivotal role in this network, and the F7 node (leadership decision factor) remains the largest node. B1 node (internal operating factor) and E3 node (Interest Rate risk factor) showed stronger infectivity than other nodes. The internal operational risk can be controlled by the internal mature management system of the financial enterprise. Only the interest rate risk needs to be regulated by the main body of supervision in the macro economy. However, the interest rate of a country affects the economic level of the whole market and is inevitable.

The $\overleftarrow{P(k_i)}$ of each node in Fig.2(c) represents the sensitivity of a node to attack. The actual meaning is the number of risks transmitted directly to the node after the occurrence of systemic risk. The order of six risk classes from big to small is: node F (Operational Strategy), E node (Business risk category), C node (Legal risk category), D node (Credit risk category), B node (Operational risk category) and A node (Technical risk category). Compared with Fig. 4, the network is more specific to the vulnerability and vulnerability of each node. The risk within the Internet enterprise remains the highest degree of connectivity It shows that Internet companies are more sensitive to risks and are more susceptible to other risks, thus triggering new risks.

From the practical point of view, the leadership decision making node, as the most important node, is not only the access point of many factors, but also the point of actively radiating links to other risk factors, which is due to the Internet Finance platform, whether it is in the development of business. The signing of contracts, the choice of financing or, for example, the use of funds, the management of enterprises, the use of user information is all decided by the corporate leadership. On the contrary, the change of laws and regulations, the change of market cycle and the information brought by technology also greatly influence the judgment and the choice of the future development direction made by the enterprise leaders. From a micro perspective, because the leadership represents Internet Finance enterprises, so the Internet Finance industry should be made up of one Internet Finance platform after it has been subdivided. In this way, the Internet Finance industry is not only itself a part of the network that spreads risks to the outside. At the same time, each platform is also vulnerable to the risk of other platforms in the industry contagion.

## 4. Regulatory Subnet

The network analysis of Internet Finance is a complex systematic project, which needs not only the internal factors of the Internet Finance industry, but also the regulatory model. Now, the cost of collecting regulatory information from the Internet industry is very high and difficult for regulators. However, considering that the Internet industry has obvious Matthew effect and long tail effect, the more important platform in the whole industry system can not only control the supervision cost and improve the regulation efficiency, but also select the more important platform in the whole industry system as the regulatory research object. And can play a role in controlling systemic risk [20]. This section marks the entire regulatory type as node G. The specific influencing factors under this type are: scale access, technology access, business license, exit license, supervision, capital supervision, public opinion supervision, legal perfection, supervision subject, central bank, local government. By analyzing the direct contagion ability of each specific factor to other risks, the risk transfer path network between the regulatory layer and the Internet Finance layer is constructed.

In February 2019, China carried out financial regulatory reform, and established the macro-prudential supervision bureau to strengthen macro-prudential supervision. Its regulatory bodies are supervised by the central bank, the three supervisory commissions and local authorities at two levels. Local supervision can ensure the implementation of Internet Finance business with local characteristics under macro supervision, which not only protects the stability of the whole market but also facilitates the innovation of Internet Finance. Therefore, the regulatory subject should be the core of the whole regulatory network and control the changes of all regulatory factors [35].

The entry, exit, regulation and other regulatory policies of the Internet Finance industry will directly affect the technical risk (category A), the legal risk (category C), the credit risk (category D), the business risk (category E) of the Internet Finance enterprise, enterprise risk (category F). For example, Internet Finance companies with insufficient technical depth cannot operate, Internet companies with a certain scale cannot carry out Internet Finance business, exit needs approval from relevant institutions and so on. Through the means of supervision in this respect, we can not only prevent excessive monopoly, vicious competition and market failure, but also guarantee the financial safety of those who participate in network finance investment and financing, and avoid it as far as possible. Network technology, management backward enterprises into the Internet Finance market.

At present, Internet Finance related law vacancy and regulation lags [43]. The supervision of Internet Finance by relevant departments also includes dynamic control, such as the continuous improvement of laws and regulations, the dynamic detection of the authenticity of funds and the timely control of public opinion. Through dynamic supervision, regulators will help to strengthen the supervision of risks, enhance the integrity of the industry, and thus affect the legal risk (category C), credit risk (category D), business risk (category E) of Internet Finance. Enterprise risk (category F).

At this stage, there is no capital reserve system in Internet Finance, which, on the one hand, makes banks in an unfair state of competition, on the other hand, it also makes Internet Finance lack the protection mechanism of the lender of last resort. Once there is a break in the capital chain, you'll be caught in an unpayable crisis. Therefore, capital supervision can make Internet Finance maintain a fair competitive market environment at the macro level, and provide the guarantee of lender of last resort for Internet Finance enterprises through capital supervision. The number and attributes of the specific impact factors of the regulatory type are shown in Table (2). The visual image of the supervised subnet is shown in Fig. 3(a), (b) and(c).

**Table 2**
Regulatory risk is divided into the following subdivision of risk factors, and added to the entire network construction.

| Node name | Number | Node name | Number | Node name | Number |
|---|---|---|---|---|---|
| | | Category G: Regulation | | | |
| Scale Access | G1 | Technology Access | G2 | Business License | G3 |
| Withdrawal Permit | G4 | Supervision Intensity | G5 | Fund Supervision | G6 |
| Public Opinion Supervision | G7 | The Law | G8 | Supervising Subject | G9 |
| Central Bank | G10 | Local Government | G11 | | |

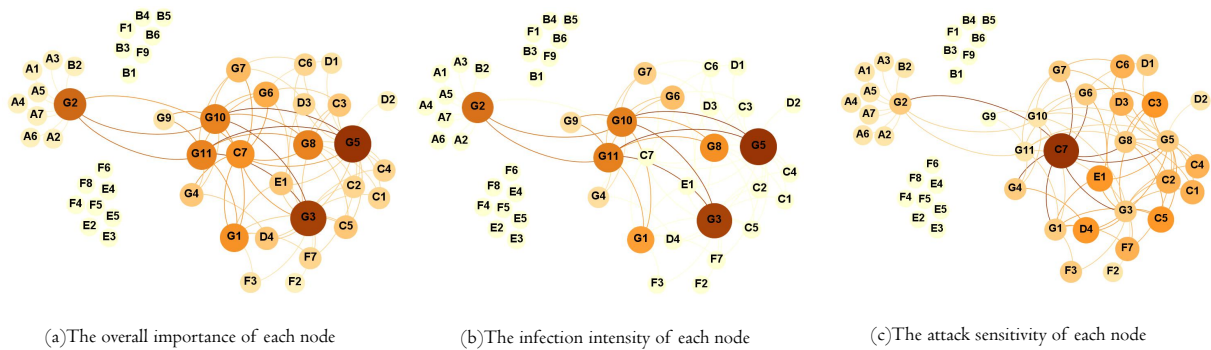

(a) The overall importance of each node  (b) The infection intensity of each node  (c) The attack sensitivity of each node

**Fig. 3** At present, the regulation is not sensitive to the change of Internet Finance, all the regulatory factors are in the low influence, B (operational risk type), E type (business risk type), category F (risk types of Internet enterprises) and so on cannot be directly linked to regulatory factors. It can be seen that existing regulatory measures cannot cover all the risks of Internet Finance, such as malicious hacking incidents, different users' preferences for products, employee relations within Internet enterprises, and so on.

## 5. Traditional Financial Subnet

The traditional financial subnet mainly describes the complex network of Internet Finance and banking. As the central node, the two have gradually formed the relationship between competition and cooperation in the development. The subnet has four branches: direct generation, indirect generation, contact infection and non-contact infection.

There is a risk generation mechanism between Internet Finance and traditional finance. Through the Internet lending business, the Internet fund business, and the Internet payment business, the traditional banks have been impacted by the successful crowding out of commercial banks' assets, liabilities and intermediary business, resulting in higher operating costs and lower profit levels of the banks. The loss of deposits and the increase in leverage. The service side of Internet Finance for users has changed the level of money supply and demand and interest rate, which has an impact on the intermediary variables of macro economy. This led to the expansion of bank credit, the decline in monetary demand, and the weakening of the effectiveness of macroeconomic controls, leading to a systemic crisis in the banking sector [11, 44].

There is a risk contagion mechanism between Internet Finance and traditional finance. Internet Finance through the channels of capital exchange generated by cooperation with commercial banks, through the Internet Finance industry platform, based on bank electronic accounts, deposit accounts, depository accounts, reserve accounts, and so on. The whole industry risk, service object risk, legal supervision risk and technical operation risk of Internet Finance are transmitted to commercial banks. Internet Finance users continue to grow and gradually become dependent on the platform. If the payment services provided by the Internet platform are suddenly paralyzed and not repaired in time, the crisis news passes through the media channels. The herding effect may cause investors to change their psychological expectations, such as cash runs and other social events [45, 46]. The traditional financial risk type is labeled as H. The numbers and attributes of the specific risk factors are shown in Table 3. The visual image is shown in Fig. 4.

**Table 3**
In view of the research of Internet Finance, this paper divides the traditional financial risk types into the following sub-risks and adds them to the construction of the network.

| Node name | Number | Node name | Number | Node name | Number |
|---|---|---|---|---|---|
| | | Category H: Traditional Financial | | | |
| Bank Deposit | H1 | Bank Debt | H2 | Intermediate Business of Bank | H3 |
| Bank Server | H4 | Insurance | H5 | Fund | H6 |

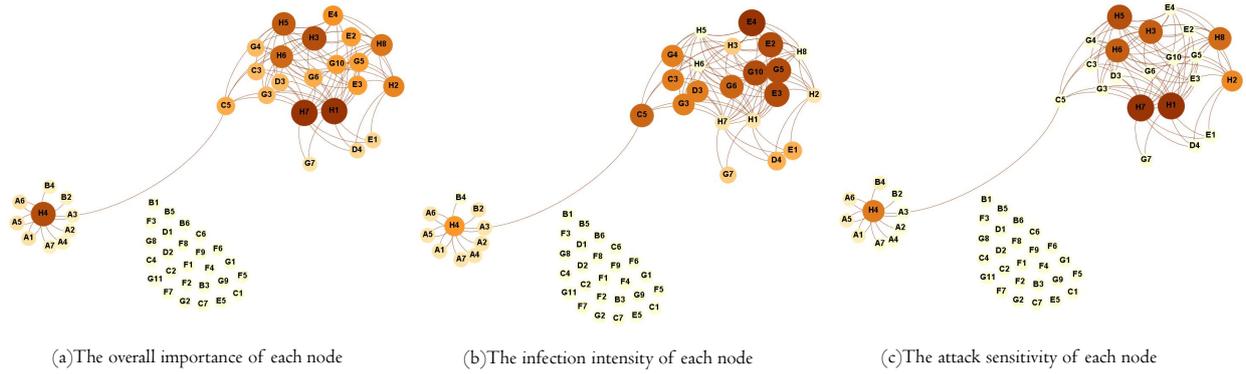

(a) The overall importance of each node    (b) The infection intensity of each node    (c) The attack sensitivity of each node

**Fig. 4** The number of nodes connected between Internet Finance and traditional finance is less than that of other subnets. The two networks mainly use business risk (category E) and credit risk (category D) as the transmission path. The most prominent is the H4 node (the pressure of internet banking service). As the final settlement link of Internet Finance transactions, banks play an important role, and the service ability of internet banking also has a direct impact on the whole Internet Finance business.

## 6. Systematic risk network

By synthesizing the above network divisions for supervision, Internet Finance, and traditional finance, it can be proved that Internet Finance risks meet the three conditions: they are composed of core outward diffusion, and each part is mutually transmitted and affected. Each part has its own unique nature. Thus, it can be seen that the composition of Internet Finance risk is actually a multi-level, outward-spreading structure, which has the possibility of inducing systemic risk. In addition, in the actual analysis of the risk situation of Internet Finance, we also have to consider the risks brought by users, the risks brought by domestic environment, and the risks brought by the main types of Internet Finance services. So this article adds some user factor types (category I), domestic and foreign environment types (category J), Internet Finance business types (category K). Specific factors and their attributes are shown in Table 4. According to the multi-layer network theory of systemic risk [30, 31], this paper selects the three-layer subnet constructed for splicing to obtain the network of Internet Finance systemic risk, so as to better observe the node connection, risk contagion path and system structure in the Internet Finance system. Because of the overlap of nodes and nodes between the three layer networks, the three layer networks must be combined into a new complex network. Similarly, the size of the $P(k_i)$, $P(k_i)^{\leftarrow}$, $P(k_i)^{\rightarrow}$ data is based on Fruch Visual images of the Terman Reingold model, as shown in Fig. 5 (a), (b), (c).

**Table 4**
User factors, domestic and foreign environment, Internet Finance services node.

| Node name | Number | Node name | Number | Node name | Number |
|---|---|---|---|---|---|
| Category I: User factors | | | | | |
| Risk Preference | I1 | Wage Level | I2 | Level of Consumption | I3 |
| Investment Experience | I4 | Risk Assumption Level | I5 | Asset Level | I6 |
| Bank Credit Evaluation | I7 | Tax Degree | I8 | Age | I9 |
| Endemic Distribution | I10 | Local Economy | I11 | | |
| Category J: Domestic and Foreign Environment | | | | | |
| Domestic Innovation | J1 | Foreign Innovation | J2 | Foreign Trade Level | J3 |
| Domestic Industry Level | J4 | Exchange of Know-how | J5 | Domestic and Foreign Capital Relations | J6 |
| Category K: Internet Finance Services | | | | | |
| Payment Service | K1 | Insurance Services | K2 | Investment Services | K3 |
| Credit Services | K4 | Money Fund | K5 | | |

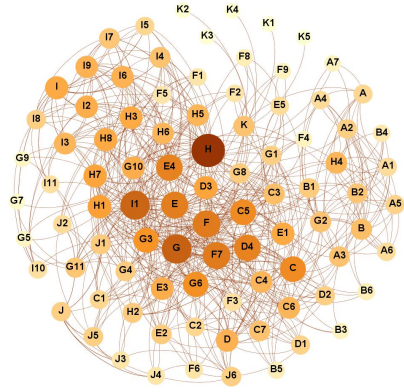

(a) The overall importance of each node

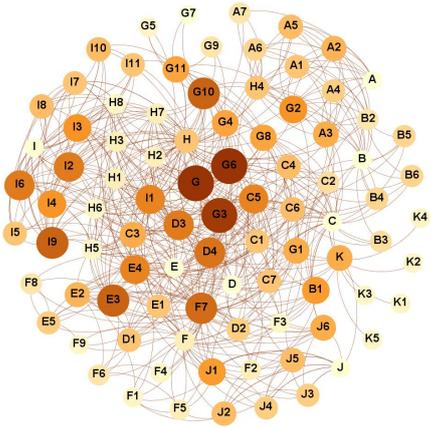

(b) The infection intensity of each node

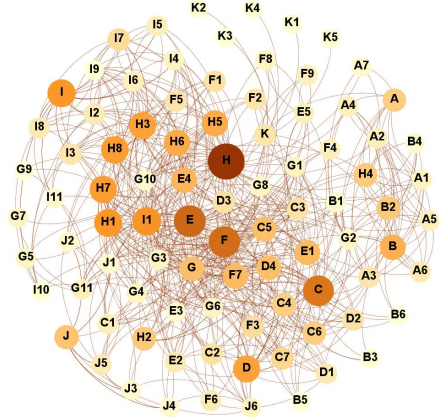

(c) The attack sensitivity of each node

**Fig. 5** Network structure Diagram of Internet Finance risk system with three subnets.

From the view of the correlation structure of the Internet Finance risk system, many elements will change with the change of time and space. For example, the continuous improvement of the regulatory system, the extinction of the old risks and the emergence of new risks. The emergence of new Internet Finance services, etc. Therefore, the system is dynamic and a complex multivariable system. In order to maintain a stable state during the continuous evolution of the system, it is necessary to balance the structure of all the components of the system with the external environment at t time. When the external environment appears random sudden factors and impact on the system, the system will usually adapt itself, this article will be the Internet Finance system Expressed as:

$$IFS(\delta, t) = [C(\delta, t), E(\delta, t), S(\delta, t)] \tag{4}$$

$IFS(\delta, t)$ stands for an Internet Finance system at a certain level, $\delta$ for the system, $t$ for the measuring time, $C$ for the Internet Finance subnet, $E$ for the regulatory subnet, and $S$ for the traditional financial subnet. In fact, the Internet Finance system adjusts its risk through feedback mechanism to realize the smooth integration of the three subnets.

The characteristics of systemic risk contagion in Internet Finance are shown in Fig. 6.

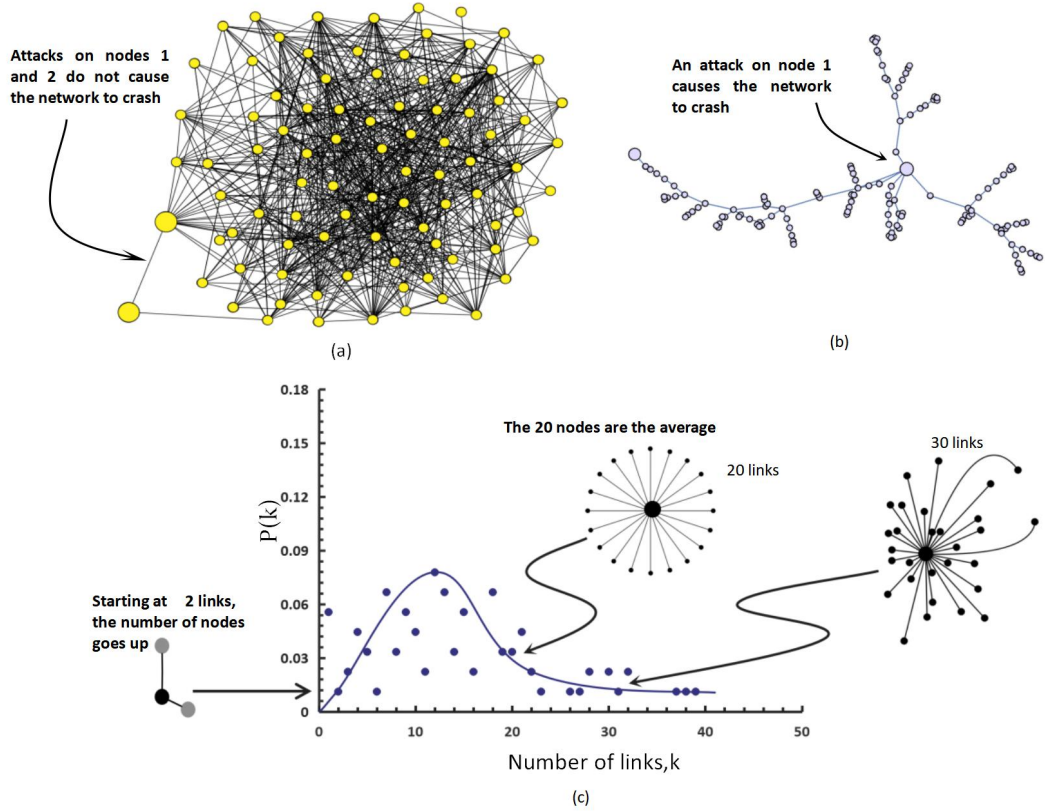

**Fig. 6** (a) and (b) show that in the whole Internet Finance complex network, the random destruction of nodes outside the system will not lead to the collapse of the whole system, but there are several core nodes in the Internet Finance risk system. These core important nodes such as G (regulatory control), H (traditional financial risk), F (operation and strategy) and other. (c) are the degree distribution of each risk factor. The image shows that the network is not a significant scale-free network. Starting with the second link, the number of nodes begins to rise 20 links are the average of the node links for the entire risk system, before which the image is presented Normal distribution, after which the image presents a power law distribution.

## 7. Choice of Investors: Traditional Finance or Internet Finance?

In the entire financial market, there is a business competition between traditional finance and Internet finance. Internet finance provides a convenient financing platform for both the supply and demand sides of funds, and builds a bridge to improve the efficiency of asset allocation. However, compared with the long-established traditional finance, the future of Internet finance, which is an emerging industry, is full of uncertainty. This uncertainty and the external environment formed a joint impact, not only led to regulatory difficulties, but also led to the investors' decision-making mistakes[46][47]. In the face of convenient services provided by Internet finance, investors often care about risk issues. It can be seen that risks and benefits determine the competitiveness of traditional finance and Internet finance. In financial markets with high uncertainties, the trade-off between maximizing expected return and minimizing the risk is one of the main challenges in modeling and decision making[48][49].

Therefore, we borrowed the mechanism of capital flow and quantitatively studied the internal relations between investors, capital demanders, traditional finance, and Internet finance in the process of investment and financing, aiming to abstract the competition problem of the financial industry into the equilibrium problem of the network[50][51]. Through the network analysis method, the trade-off between profit maximization and risk minimization is judged, and the competition relationship and risk degree between Internet finance and traditional finance are estimated based on the investor's preference[52][53][54].

### 7.1 Investor behavior analysis

Most Internet companies pay great attention to business models and business innovations. Focusing on big data processing and user behavior analysis, they can often achieve accurate business promotion in vertical segments. The Internet financial platform attracts potential investors in Internet applications such as search engines, social networking, and entertainment by using bonus payment and redemption points.

Defining the amount of funds the investor h has is $q_{hi}^1$, the fund owner h invests the funds in an internet financial platform i, so there are HI links linking the investors and the internet financial platform. The flow of funds on each link $q_{hi}^1$ is summed to form a vector $Q^1 \in R_+^{HI}$.

At this point investors can have the following options:

(1) Investors transfer funds to the Internet financial platform, and the latter directly conducts investment activities, such as Internet P2P credit.

The Internet financial platform i directly lends the funds to the capital demander k, thus generating IK links between each of them. The flow of funds on each link is $q_{ik}^2$, aggregating to form a vector $Q^4 \in R_+^{IK}$.

(2) Investors transfer funds to traditional financial platforms, such as bank time deposits and other investment activities.

Investor h chooses to invest his funds in a traditional financial platform j. Therefore, a total of HJ links exist among investors and traditional financial platforms, and the flow of funds on each link is $q_{hj}^1$, aggregating to form a vector $Q^2 \in R_+^{HJ}$.

(3) Investors transfer funds to an internet financial platform. At this time, the Internet financial platform plays an intermediate role, and continues to pool funds into traditional financial platforms and reach the specific capital demand market through traditional finance.

The Internet financial platform i transfers funds of $q_{ij}^2$ to traditional financial platform j for investment. A total of IJ links exist among Internet finance and traditional finance, and the flow of funds on each link is $q_{ij}^2$, aggregating to form a vector $Q^3 \in R_+^{IJ}$. Therefore, the summary vector of all Internet financial platforms $Q^5$ should be the summation of $Q^2 \in R_+^{HJ}$ and $Q^3 \in R_+^{IJ}$.

### 7.2 Capital flow and risk function

Since financial instruments such as "leverage investment" are not yet mature in the Internet finance industry, with no specific laws and regulations on them, this paper assumes the funds used for foreign investment in the Internet financial platform is raised from market. For each Internet financial platform, the amount of funds flowing out cannot exceed the amount of funds flowing in.

$$\sum_{j=1}^{J} q_{ij}^2 + \sum_{k=1}^{K} q_{ik}^2 \leq \sum_{h=1}^{H} q_{hi}^1 \quad \forall i \tag{5}$$

The definition of $\eta_{ij}^2$ indicates the existence of the relationship between the financial business platform i and the traditional financial platform j, and aggregates the formation of the vector $H^3 \in R_+^{IJ}$; $\eta_{ik}^2$ represents the capital flow relationship between the Internet financial platform i and the capital demander k, and aggregates to vector $H^4 \in R_+^{IK}$. Assume that the strength of these network relationships is in the interval [0, 1], 0 means that the network relationship between the two does not exist, and 1 represents the closest social network relationship between the two.

The cost function for the relationship between the Internet financial platform i and the traditional financial platform j is defined as $f_{ij}^2 = f_{ij}^2(\eta_{ij}^2) \quad \forall i,j$. The cost function for the relationship between the Internet financial platform and the capital demanders is defined as $f_{ik}^2 = f_{ik}^2(\eta_{ik}^2) \quad \forall i,k$. The cost function for the relationship between investors and internet finance platforms is defined as $\hat{f}_{hi}^1 = \hat{f}_{hi}^1(\eta_{hi}^1) \quad \forall h,i$.

In order to facilitate the optimization of the solution, the horizontal function of the relationship between the Internet financial platform i and the traditional financial platform j is defined as $v_{ij}^2 = v_{ij}^2(\eta_{ij}^2) \quad \forall i,j$. The horizontal function of the relationship between the Internet financial platform and the capital demanders is defined as $v_{ik}^2 = v_{ik}^2(\eta_{ik}^2) \quad \forall i,k$. The horizontal function function for the relationship between the fund owner and the internet financial platform is defined as $\hat{v}_{hi}^1 = \hat{v}_{hi}^1(\eta_{hi}^1) \quad \forall h,i$.

Define $c_{ij}^2$ as the transaction cost function between the Internet financial platform i and the traditional financial platform j (the part of the Internet financial platform). It is assumed that transaction costs are determined based on each other's trading volume and relationship level: $c_{ij}^2 = c_{ij}^2(q_{ij}^2, \eta_{ij}^2) \quad \forall i,j$. Transaction cost function between the Internet financial platform i and the capital demander k is defined as $c_{ik}^2 = c_{ik}^2(q_{ik}^2, \eta_{ik}^2) \quad \forall i,k$. Transaction cost function between the fund owner h and the internet financial platform i is defined as $\hat{c}_{hi}^1 = \hat{c}_{hi}^1(q_{hi}^1, \eta_{hi}^1) \quad \forall h,i$.

Define $c_i$ as the conversion cost function of the Internet financial platform i, including the conversion cost of the funds invested by the recipient of the funds into the traditional financial platform and related financial products required by the capital demanders. Assume that the conversion cost function is based on the amount of money $\sum_{h=1}^{H} q_{hi}^1$ flowing into the Internet financial platform i: $c_i = c_i(Q^1) \quad \forall i$.

Define $e_{ij}^2$ as an operational risk function between the Internet financial platform i and the traditional financial platform j. The operational risk function between the Internet financial platform i and the capital demander k is defined as $e_{ik}^2 = a_i^2 E_{ik}^2(q_{ik}^2) - b_i^2 T_{ik}^2(q_{ik}^2) \quad \forall i,k$.

It is assumed that the operating profit function represents the excess return obtained due to the operational skills of the Internet financial platform, and the technical penalty function represents the loss caused by the failure of the Internet financial platform equipment, which is borne by the Internet financial platform itself. For the operational skill coefficient $a_i^2$ of the Internet financial platform, it is only relevant to the Internet financial platform side. The component $E_{ij}^2$ of the operational profit function are determined based on the volume of each other's transactions. For the technical risk factor $b_i^2$ of the Internet financial platform, this coefficient is only relevant to the Internet financial platform. The component $T_{ij}^2$ of the technical penalty function are determined based on each other's trading volume: $e_{ij}^2 = a_i^2 E_{ij}^2(q_{ij}^2) - b_i^2 T_{ij}^2(q_{ij}^2) \quad \forall i,j$.

Define $g_{ij}^2$ as the credit penalty function between the Internet financial platform i and the traditional financial platform j (the part of the Internet financial platform that defaults, and thus is undertaken by it). Assume that the credit penalty function has a relationship with both parties to the transaction, and is determined based on each other's transaction volume and relationship level: $g_{ij}^2 = g_{ij}^2(q_{ij}^2, \eta_{ij}^2) \quad \forall i,j$. The credit penalty function between the Internet financial platform i and the capital demander k is defined as $g_{ik}^2 = g_{ik}^2(q_{ik}^2, \eta_{ik}^2) \quad \forall i,k$. The credit penalty function between the fund owner h and the internet financial platform i is defined as $\hat{g}_{hi}^1 = \hat{g}_{hi}^1(q_{hi}^1, \eta_{hi}^1) \quad \forall h,i$. Define $r_{ij}^2$ as risk function between the Internet financial platform i and the traditional financial platform j. Assume that the risk function is not only related to the volume of each other's transactions, but also to the level of relationship with each other. A rise in the level of relationship leads to a higher level of trust, at the same time reducing the uncertainty of the transaction and thus reduces the risk. Meanwhile, a higher level of relationship will increase the competitiveness of decision makers in the market: $r_{ij}^2 = r_{ij}^2(q_{ij}^2, \eta_{ij}^2) \quad \forall i,j$.

Define the risk function between the Internet financial platform i and the capital demander k as $r_{ik}^2 = r_{ik}^2(q_{ik}^2, \eta_{ik}^2) \quad \forall i,k$. The risk function between the fund owner h and the internet financial platform i is defined as $\hat{r}_{hi}^1 = \hat{r}_{hi}^1(q_{hi}^1, \eta_{hi}^1) \quad \forall h,i$.

It is assumed that the above relationship cost function, relationship level function, transaction cost function, operation risk function, credit penalty function, and risk function are both convex functions and continuously differentiable.

### 7.3 Revenue, risk and user stickiness

For the Internet financial platform, there is a need for financial institutions to balance the risks and benefits, and it is also extremely concerned to maintain contact with users, thereby increasing the user's stickiness of products. Therefore, it has the characteristics of maximizing profit, minimizing risk and maximizing user stickiness.

**Goal 1**: Maximize net income

Define $\rho_{ij}^2$ as the price of funds transferred from the Internet financial platform i to traditional finance j, which is handled by traditional finance.

Define $\rho_{ik}^2$ as the price of funds directly borrowed from the Internet financial platform i to the capital demander k.

Then the problem of maximizing net income of the Internet financial platform i is expressed as:

$$\max z_{1i}^2 = \sum_{j=1}^{J}\left(\rho_{ij}^2 q_{ij}^2 + e_{ij}^2\left(q_{ij}^2\right) - f_{ij}^2\left(\eta_{ij}^2\right) - c_{ij}^2\left(q_{ij}^2,\eta_{ij}^2\right) - g_{ij}^2\left(q_{ij}^2,\eta_{ij}^2\right)\right)$$
$$+ \sum_{k=1}^{K}\left(\rho_{ik}^2 q_{ik}^2 + e_{ik}^2\left(q_{ik}^2\right) - f_{ik}^2\left(\eta_{ik}^2\right) - c_{ik}^2\left(q_{ik}^2,\eta_{ik}^2\right) - g_{ik}^2\left(q_{ik}^2,\eta_{ik}^2\right)\right)$$
$$- c_i\left(Q^1\right) - \sum_{h=1}^{H}\left(\rho_{hi}^1 q_{hi}^1 + \hat{f}_{hi}^1\left(q_{hi}^1\right) + \hat{c}_{hi}^1\left(q_{hi}^1,\eta_{hi}^1\right) + \hat{g}_{hi}^1\left(q_{hi}^1,\eta_{hi}^1\right)\right)$$
. (6)

Also meet the conditions $q_{hi}^1 \geq 0, q_{ij}^2 \geq 0, q_{ik}^2 \geq 0 \quad \forall h,i,j,k$ and $0 \leq \eta_{hi}^1 \leq 1, 0 \leq \eta_{ij}^2 \leq 1, 0 \leq \eta_{ik}^2 \leq 1 \quad \forall h,i,j,k$ as well.

**Goal 2**: Minimize risk
The risk minimization problem of the Internet financial platform i is expressed as:
$$\min z_{2i}^2 = \sum_{j=1}^{J} r_{ij}^2\left(q_{ij}^2,\eta_{ij}^2\right) + \sum_{k=1}^{K} r_{ik}^2\left(q_{ik}^2,\eta_{ik}^2\right) + \sum_{h=1}^{H} \hat{r}_{hi}^1\left(q_{hi}^1,\eta_{hi}^1\right)$$
. (7)

Also meet the conditions $q_{hi}^1 \geq 0, q_{ij}^2 \geq 0, q_{ik}^2 \geq 0 \quad \forall h,i,j,k$ and $0 \leq \eta_{hi}^1 \leq 1, 0 \leq \eta_{ij}^2 \leq 1, 0 \leq \eta_{ik}^2 \leq 1 \quad \forall h,i,j,k$ as well.

**Goal 3**: Maximize user stickiness
The Internet finance platform also attempts to establish and maintain the closest social relationship with other decision makers on the network. This question is expressed as:
$$\max z_{3i}^2 = \sum_{j=1}^{J} v_{ij}^2\left(\eta_{ij}^2\right) + \sum_{k=1}^{K} v_{ik}^2\left(\eta_{ik}^2\right) + \sum_{h=1}^{H} \hat{v}_{hi}^1\left(\eta_{hi}^1\right)$$
. (8)

Also meet the condition $0 \leq \eta_{hi}^1 \leq 1, 0 \leq \eta_{ij}^2 \leq 1, 0 \leq \eta_{ik}^2 \leq 1 \quad \forall h,i,j,k$.

### 7.4 Network balance analysis

For the multi-dimensional goal of the Internet financial platform i, define $U_i$ as multi-objective decision function, $\alpha_i$ as non-negative risk weight and $\beta_i$ as non-negative relationship weight.

$$\max U_i = \sum_{j=1}^{J}\left(\rho_{ij}^2 q_{ij}^2 + e_{ij}^2\left(q_{ij}^2\right) - f_{ij}^2\left(\eta_{ij}^2\right) - c_{ij}^2\left(q_{ij}^2,\eta_{ij}^2\right) - g_{ij}^2\left(q_{ij}^2,\eta_{ij}^2\right)\right)$$
$$+ \sum_{k=1}^{K}\left(\rho_{ik}^2 q_{ik}^2 + e_{ik}^2\left(q_{ik}^2\right) - f_{ik}^2\left(\eta_{ik}^2\right) - c_{ik}^2\left(q_{ik}^2,\eta_{ik}^2\right) - g_{ik}^2\left(q_{ik}^2,\eta_{ik}^2\right)\right)$$
$$- c_i\left(Q^1\right) - \sum_{h=1}^{H}\left(\rho_{hi}^1 q_{hi}^1 + \hat{f}_{hi}^1\left(\eta_{hi}^1\right) + \hat{c}_{hi}^1\left(q_{hi}^1,\eta_{hi}^1\right) + \hat{g}_{hi}^1\left(q_{hi}^1,\eta_{hi}^1\right)\right)$$
$$- \alpha_i\left(\sum_{j=1}^{J} r_{ij}^2\left(q_{ij}^2,\eta_{ij}^2\right) + \sum_{k=1}^{K} r_{ik}^2\left(q_{ik}^2,\eta_{ik}^2\right) + \sum_{i=1}^{I} \hat{r}_{hi}^1\left(q_{hi}^1,\eta_{hi}^1\right)\right)$$
$$+ \beta_i\left(\sum_{j=1}^{J} v_{ij}^2\left(\eta_{ij}^2\right) + \sum_{k=1}^{K} v_{ik}^2\left(\eta_{ik}^2\right) + \sum_{i=1}^{I} \hat{v}_{hi}^1\left(\eta_{hi}^1\right)\right)$$
. (9)

with the conditions $q_{hi}^1 \geq 0, q_{ij}^2 \geq 0, q_{ik}^2 \geq 0 \quad \forall h,i,j,k$, $0 \leq \eta_{hi}^1 \leq 1, 0 \leq \eta_{ij}^2 \leq 1, 0 \leq \eta_{ik}^2 \leq 1 \quad \forall h,i,j,k$, $\sum_{j=1}^{J} q_{ij}^2 + \sum_{k=1}^{K} q_{ik}^2 \leq \sum_{h=1}^{H} q_{hi}^1 \quad \forall i$. The same is true with (28), which is a strict concave function, therefore we could use variational inequality to represent the equilibrium conditions that satisfy all Internet financial platforms:

$$\sum_{i=1}^{I}\sum_{j=1}^{J}\left[\alpha_i\frac{\partial r_{ij}^2\left(q_{ij}^{2*},\eta_{ij}^{2*}\right)}{\partial q_{ij}^2}+\frac{\partial c_{ij}^2\left(q_{ij}^{2*},\eta_{ij}^{2*}\right)}{\partial q_{ij}^2}+\frac{\partial g_{ij}^2\left(q_{ij}^{2*},\eta_{ij}^{2*}\right)}{\partial q_{ij}^2}-\frac{\partial e_{ij}^2\left(q_{ij}^{2*}\right)}{\partial q_{ij}^2}-\rho_{ij}^*+\gamma_i^*\right]\times\left(q_{ij}^2-q_{ij}^{2*}\right)$$

$$+\sum_{i=1}^{I}\sum_{k=1}^{K}\left[\alpha_i\frac{\partial r_{ik}^2\left(q_{ik}^{2*},\eta_{ik}^{2*}\right)}{\partial q_{ik}^2}+\frac{\partial c_{ik}^2\left(q_{ik}^{2*},\eta_{ik}^{2*}\right)}{\partial q_{ik}^2}+\frac{\partial g_{ik}^2\left(q_{ik}^{2*},\eta_{ik}^{2*}\right)}{\partial q_{ik}^2}-\frac{\partial e_{ik}^2\left(q_{ik}^{2*}\right)}{\partial q_{ik}^2}-\rho_{ik}^{2*}+\gamma_i^*\right]\times\left(q_{ik}^2-q_{ik}^{2*}\right)$$

$$+\sum_{h=1}^{H}\sum_{i=1}^{I}\left[\alpha_i\frac{\partial \hat{r}_{hi}^1\left(q_{hi}^{1*},\eta_{hi}^{1*}\right)}{\partial \eta_{hi}^1}+\frac{\partial c_i\left(Q^{1*}\right)}{\partial q_{hi}^1}+\rho_{hi}^{1*}+\frac{\partial c_{hi}^1\left(q_{hi}^{1*},\eta_{hi}^{1*}\right)}{\partial q_{hi}^1}+\frac{\partial g_{hi}^1\left(q_{hi}^{1*},\eta_{hi}^{1*}\right)}{\partial q_{hi}^1}-\gamma_i^*\right]\times\left(q_{hi}^1-q_{hi}^{1*}\right)$$

$$+\sum_{i=1}^{I}\sum_{j=1}^{J}\left[\alpha_i\frac{\partial r_{ij}^2\left(q_{ij}^{2*},\eta_{ij}^{2*}\right)}{\partial \eta_{ij}^2}+\frac{\partial f_{ij}^2\left(\eta_{ij}^{2*}\right)}{\partial \eta_{ij}^2}+\frac{\partial c_{ij}^2\left(q_{ij}^{2*},\eta_{ij}^{2*}\right)}{\partial \eta_{ij}^2}+\frac{\partial g_{ij}^2\left(q_{ij}^{2*},\eta_{ij}^{2*}\right)}{\partial \eta_{ij}^2}-\beta_i\frac{\partial v_{ij}^2\left(\eta_{ij}^{2*}\right)}{\partial \eta_{ij}^2}\right]\times\left(\eta_{ij}^2-\eta_{ij}^{2*}\right)$$

(10).

$$+\sum_{i=1}^{I}\sum_{k=1}^{K}\left[\alpha_i\frac{\partial r_{ik}^2\left(q_{ik}^{2*},\eta_{ik}^{2*}\right)}{\partial \eta_{ik}^2}+\frac{\partial f_{ik}^2\left(\eta_{ik}^{2*}\right)}{\partial \eta_{ik}^2}+\frac{\partial c_{ik}^2\left(q_{ik}^{2*},\eta_{ik}^{2*}\right)}{\partial \eta_{ik}^2}+\frac{\partial g_{ik}^2\left(q_{ik}^{2*},\eta_{ik}^{2*}\right)}{\partial \eta_{ik}^2}-\beta_i\frac{\partial v_{ik}^2\left(\eta_{ik}^{2*}\right)}{\partial \eta_{ik}^2}\right]\times\left(\eta_{ik}^2-\eta_{ik}^{2*}\right)$$

$$+\sum_{h=1}^{H}\sum_{i=1}^{I}\left[\alpha_i\frac{\partial \hat{r}_{hi}^1\left(q_{hi}^{1*},\eta_{hi}^{1*}\right)}{\partial \eta_{hi}^1}+\frac{\partial \hat{f}_{hi}^1\left(\eta_{hi}^{1*}\right)}{\partial \eta_{hi}^1}+\frac{\partial c_{hi}^1\left(q_{hi}^{1*},\eta_{hi}^{1*}\right)}{\partial \eta_{hi}^1}+\frac{\partial g_{hi}^1\left(q_{hi}^{1*},\eta_{hi}^{1*}\right)}{\partial \eta_{hi}^1}-\beta_i\frac{\partial v_{hi}^1\left(\eta_{hi}^{1*}\right)}{\partial \eta_{hi}^1}\right]\times\left(\eta_{hi}^1-\eta_{hi}^{1*}\right)$$

$$+\sum_{i=1}^{I}\left(\sum_{h=1}^{H}q_{hi}^{1*}-\sum_{j=1}^{J}q_{ij}^{2*}-\sum_{k=1}^{K}q_{ik}^{2*}\right)\times\left(\gamma_i-\gamma_i^*\right)\geq 0,$$

$$\forall\left(Q^1,Q^3,Q^4,H^1,H^3,H^4,\Gamma\right)\in K^2$$

$$K^2\equiv\left\{\left(Q^1,Q^3,Q^4,H^1,H^3,H^4,\Gamma\right)\mid q_{hi}^1\geq 0, q_{ij}^2\geq 0, q_{ik}^2\geq 0, 0\leq\eta_{hi}^1\leq 1, 0\leq\eta_{ij}^2\leq 1, 0\leq\eta_{ik}^2\leq 1, \gamma_i\geq 0, \forall h,i,j,k\right\}$$

The variational inequality is used to represent the equilibrium conditions that satisfy all investors as follows:

$$\sum_{h=1}^{H}\sum_{i=1}^{I}\left[\alpha_h\frac{\partial r_{hi}^1\left(q_{hi}^{1*},\eta_{hi}^{1*}\right)}{\partial q_{hi}^1}+\frac{\partial c_{hi}^1\left(q_{hi}^{1*},\eta_{hi}^{1*}\right)}{\partial q_{hi}^1}+\frac{\partial g_{hi}^1\left(q_{hi}^{1*},\eta_{hi}^{1*}\right)}{\partial q_{hi}^1}-\rho_{hi}^{1*}\right]\times\left(q_{hi}^1-q_{hi}^{1*}\right)$$

$$+\sum_{h=1}^{H}\sum_{j=1}^{J}\left[\alpha_h\frac{\partial r_{hj}^1\left(q_{hj}^{1*},\eta_{hj}^{1*}\right)}{\partial q_{hj}^1}+\frac{\partial c_{hj}^1\left(q_{hj}^{1*},\eta_{hj}^{1*}\right)}{\partial q_{hj}^1}+\frac{\partial g_{hj}^1\left(q_{hj}^{1*},\eta_{hj}^{1*}\right)}{\partial q_{hj}^1}-\rho_{hj}^{1*}\right]\times\left(q_{hj}^1-q_{hj}^{1*}\right)$$

$$+\sum_{h=1}^{H}\sum_{i=1}^{I}\left[\alpha_h\frac{\partial r_{hi}^1\left(q_{hi}^{1*},\eta_{hi}^{1*}\right)}{\partial \eta_{hi}^1}+\frac{\partial f_{hi}^1\left(\eta_{hi}^{1*}\right)}{\partial \eta_{hi}^1}+\frac{\partial c_{hi}^1\left(q_{hi}^{1*},\eta_{hi}^{1*}\right)}{\partial \eta_{hi}^1}+\frac{\partial g_{hi}^1\left(q_{hi}^{1*},\eta_{hi}^{1*}\right)}{\partial \eta_{hi}^1}-\beta_h\frac{\partial v_{hi}^1\left(\eta_{hi}^{1*}\right)}{\partial \eta_{hi}^1}\right]\times\left(\eta_{hi}^1-\eta_{hi}^{1*}\right)$$

(11).

$$+\sum_{h=1}^{H}\sum_{j=1}^{J}\left[\alpha_h\frac{\partial r_{hj}^1\left(q_{hj}^{1*},\eta_{hj}^{1*}\right)}{\partial \eta_{hj}^1}+\frac{\partial f_{hj}^1\left(\eta_{hj}^{1*}\right)}{\partial \eta_{hj}^1}+\frac{\partial c_{hj}^1\left(q_{hj}^{1*},\eta_{hj}^{1*}\right)}{\partial \eta_{hj}^1}+\frac{\partial g_{hj}^1\left(q_{hj}^{1*},\eta_{hj}^{1*}\right)}{\partial \eta_{hj}^1}-\beta_h\frac{\partial v_{hj}^1\left(\eta_{hj}^{1*}\right)}{\partial \eta_{hj}^1}\right]\times\left(\eta_{hj}^1-\eta_{hj}^{1*}\right)\geq 0$$

$$\forall\left(Q^1,Q^2,H^1,H^2\right)\in K^1$$

$$K^1\equiv\left\{\left(Q^1,Q^2,H^1,H^2\right)\mid q_{hi}^1\geq 0, q_{hj}^1\geq 0, 0\leq\eta_{hi}^1\leq 1, 0\leq\eta_{hj}^1\leq 1, 且满足式\sum_{i=1}^{I}q_{hi}^1+\sum_{j=1}^{J}q_{hj}^1\leq S_h, \forall h,i,j\right\}$$

$$K^1=\left\{\left(Q^1,Q^2,H^1,H^2\right)\mid q_{hi}^1\geq 0, q_{hj}^1\geq 0, 0\leq\eta_{hi}^1\leq 1, 0\leq\eta_{hj}^1\leq 1\right\}$$

The variational inequality is used to represent the equilibrium conditions for meeting traditional financial platforms as follows:

$$\sum_{j=1}^{J}\sum_{k=1}^{K}\left[\alpha_j \frac{\partial r_{jk}^3\left(q_{jk}^{3*},\eta_{jk}^{3*}\right)}{\partial q_{jk}^3} + \frac{\partial c_{jk}^3\left(q_{jk}^{3*},\eta_{jk}^{3*}\right)}{\partial q_{jk}^3} + \frac{\partial g_{jk}^3\left(q_{jk}^{3*},\eta_{jk}^{3*}\right)}{\partial q_{jk}^3} - \frac{\partial e_{jk}^3\left(q_{jk}^{3*}\right)}{\partial q_{jk}^3} - \rho_{jk}^{3*} + \gamma_j^*\right] \times \left(q_{jk}^3 - q_{jk}^{3*}\right)$$

$$+\sum_{h=1}^{H}\sum_{j=1}^{J}\left[\alpha_j \frac{\partial \hat{r}_{hj}^1\left(q_{hj}^{1*},\eta_{hj}^{1*}\right)}{\partial q_{hj}^1} + \frac{\partial c_j\left(Q^{2*}\right)}{\partial q_{hj}^1} + \rho_{hj}^{1*} + \frac{\partial \hat{c}_{hj}^1\left(q_{hj}^{1*},\eta_{hj}^{1*}\right)}{\partial q_{hj}^1} + \frac{\partial \hat{g}_{hj}^1\left(q_{hj}^{1*},\eta_{hj}^{1*}\right)}{\partial q_{hj}^1} - \gamma_j^*\right] \times \left(q_{hj}^1 - q_{hj}^{1*}\right)$$

$$+\sum_{i=1}^{I}\sum_{j=1}^{J}\left[\alpha_j \frac{\partial \hat{r}_{ij}^2\left(q_{ij}^{2*},\eta_{ij}^{2*}\right)}{\partial q_{ij}^2} + \frac{\partial c_j\left(Q^{3*}\right)}{\partial q_{ij}^2} + \rho_{ij}^{2*} + \frac{\partial \hat{c}_{ij}^2\left(q_{ij}^{2*},\eta_{ij}^{2*}\right)}{\partial q_{ij}^2} + \frac{\partial \hat{g}_{ij}^2\left(q_{ij}^{2*},\eta_{ij}^{2*}\right)}{\partial q_{ij}^2} - \gamma_j^*\right] \times \left(q_{ij}^2 - q_{ij}^{2*}\right)$$

$$+\sum_{j=1}^{J}\sum_{k=1}^{K}\left[\alpha_j \frac{\partial r_{jk}^3\left(q_{jk}^{3*},\eta_{jk}^{3*}\right)}{\partial \eta_{jk}^3} + \frac{\partial f_{jk}^3\left(\eta_{jk}^{3*}\right)}{\partial \eta_{jk}^3} + \frac{\partial c_{jk}^3\left(q_{jk}^{3*},\eta_{jk}^{3*}\right)}{\partial \eta_{jk}^3} + \frac{\partial g_{jk}^3\left(q_{jk}^{3*},\eta_{jk}^{3*}\right)}{\partial \eta_{jk}^3} - \beta_j \frac{\partial v_{jk}^3\left(\eta_{jk}^{3*}\right)}{\partial \eta_{jk}^3}\right] \times \left(\eta_{jk}^3 - \eta_{jk}^{3*}\right) \qquad (12)$$

$$+\sum_{h=1}^{H}\sum_{j=1}^{J}\left[\alpha_j \frac{\partial \hat{r}_{hj}^1\left(q_{hj}^{1*},\eta_{hj}^{1*}\right)}{\partial \eta_{hj}^1} + \frac{\partial \hat{f}_{hj}^1\left(\eta_{hj}^1\right)}{\partial \eta_{hj}^1} + \frac{\partial \hat{c}_{hj}^1\left(q_{hj}^{1*},\eta_{hj}^{1*}\right)}{\partial \eta_{hj}^1} + \frac{\partial \hat{g}_{hj}^1\left(q_{hj}^{1*},\eta_{hj}^{1*}\right)}{\partial \eta_{hj}^1} - \beta_j \frac{\partial \hat{v}_{hj}^1\left(\eta_{hj}^1\right)}{\partial \eta_{hj}^1}\right] \times \left(\eta_{hj}^1 - \eta_{hj}^{1*}\right)$$

$$+\sum_{i=1}^{I}\sum_{j=1}^{J}\left[\alpha_j \frac{\partial \hat{r}_{ij}^2\left(q_{ij}^{2*},\eta_{ij}^{2*}\right)}{\partial \eta_{ij}^2} + \frac{\partial \hat{f}_{ij}^2\left(\eta_{ij}^2\right)}{\partial \eta_{ij}^2} + \frac{\partial \hat{c}_{ij}^2\left(q_{ij}^{2*},\eta_{ij}^{2*}\right)}{\partial \eta_{ij}^2} + \frac{\partial \hat{g}_{ij}^2\left(q_{ij}^{2*},\eta_{ij}^{2*}\right)}{\partial \eta_{ij}^2} - \beta_j \frac{\partial \hat{v}_{ij}^2\left(\eta_{ij}^2\right)}{\partial \eta_{ij}^2}\right] \times \left(\eta_{ij}^2 - \eta_{ij}^{2*}\right)$$

$$+\sum_{j=1}^{J}\left(\sum_{h=1}^{H}q_{hj}^{1*} + \sum_{i=1}^{I}q_{ij}^{2*} - \sum_{k=1}^{K}q_{jk}^{3*}\right)\left(\gamma_j - \gamma_j^*\right) \geq 0$$

$$\forall \left(Q^2, Q^3, Q^5, H^2, H^3, H^5, \Gamma\right) \in K^3$$

$$K^3 = \left\{\left(Q^2, Q^3, Q^5, H^2, H^3, H^5\right) \middle| q_{hi}^1 \geq 0, q_{ij}^2 \geq 0, q_{jk}^3 \geq 0, 0 \leq \eta_{hj}^1 \leq 1, 0 \leq \eta_{ij}^2 \leq 1, 0 \leq \eta_{ij}^3 \leq 1\right\}$$

The variational inequality is used to indicate that the equilibrium conditions for all fund users are as follows:

$$\sum_{i=1}^{I}\sum_{k=1}^{K}\left[\rho_{ik}^{2*} + \hat{c}_{ik}^2\left(Q^{4*}, Q^{5*}, H^{4*}, H^{5*}\right) + \hat{g}_{ik}^2\left(q_{ik}^{2*}, \eta_{ik}^{2*}\right) - \rho_k^{4*}\right] \times \left(q_{ik}^2 - q_{ik}^{2*}\right)$$

$$+\sum_{j=1}^{J}\sum_{k=1}^{K}\left[\rho_{jk}^{3*} + \hat{c}_{jk}^3\left(Q^{4*}, Q^{5*}, H^{4*}, H^{5*}\right) + \hat{g}_{jk}^3\left(q_{jk}^{3*}, \eta_{jk}^{3*}\right) - \rho_k^{4*}\right] \times \left(q_{jk}^3 - q_{jk}^{3*}\right) \qquad (13)$$

$$+\sum_{k=1}^{K}\left[\sum_{i=1}^{I}q_{ik}^{2*} + \sum_{j=1}^{J}q_{jk}^{3*} - d_k\left(\rho^{4*}\right)\right] \times \left(\rho_k^4 - \rho_k^{4*}\right) \geq 0 \quad \forall \left(Q^4, Q^5, \rho^4\right) \in R_+^{IK+JK+K}$$

There is a set of optimal $\left\{q_{hi}^{1*}, q_{hj}^{1*}, q_{ij}^{2*}, q_{ik}^{2*}, q_{jk}^{3*}, \eta_{hi}^{1*}, \eta_{hj}^{1*}, \eta_{ij}^{2*}, \eta_{ik}^{2*}, \eta_{jk}^{3*}, \gamma_i^*, \gamma_j^*, \rho_k^{4*}\right\}$ satisfying the sum of variational inequalities, so that the financial market reaches equilibrium.

Equilibrium solution existence condition: assuming that there is a positive constant M, N, R, so that

$$\alpha_h \frac{\partial r_{hi}^1\left(q_{hi}^{1*},\eta_{hi}^{1*}\right)}{\partial q_{hi}^1} + \frac{\partial c_{hi}^1\left(q_{hi}^{1*},\eta_{hi}^{1*}\right)}{\partial q_{hi}^1} + \frac{\partial g_{hi}^1\left(q_{hi}^{1*},\eta_{hi}^{1*}\right)}{\partial q_{hi}^1} + \alpha_i \frac{\partial \hat{r}_{hi}^1\left(q_{hi}^{1*},\eta_{hi}^{1*}\right)}{\partial \eta_{hi}^1} + \frac{\partial c_i\left(Q^{1*}\right)}{\partial q_{hi}^1}$$
$$+ \frac{\partial \hat{c}_{hi}^1\left(q_{hi}^{1*},\eta_{hi}^{1*}\right)}{\partial q_{hi}^1} + \frac{\partial \hat{g}_{hi}^1\left(q_{hi}^{1*},\eta_{hi}^{1*}\right)}{\partial q_{hi}^1} - \gamma_i^* \geq M \quad \forall Q^1, q_{hi}^1 \geq N, \forall h, i \tag{14}$$

$$\alpha_h \frac{\partial r_{hj}^1\left(q_{hj}^{1*},\eta_{hj}^{1*}\right)}{\partial q_{hj}^1} + \frac{\partial c_{hj}^1\left(q_{hj}^{1*},\eta_{hj}^{1*}\right)}{\partial q_{hj}^1} + \frac{\partial g_{hj}^1\left(q_{hj}^{1*},\eta_{hj}^{1*}\right)}{\partial q_{hj}^1} + \alpha_j \frac{\partial \hat{r}_{hj}^A\left(q_{hj}^{1*},\eta_{hj}^*\right)}{\partial q_{hj}^1} + \frac{\partial c_j\left(Q^{2*}\right)}{\partial q_{hj}^1}$$
$$+ \frac{\partial \hat{c}_{hj}^1\left(q_{hj}^{1*},\eta_{hj}^{1*}\right)}{\partial q_{hj}^1} + \frac{\partial \hat{g}_{hj}^1\left(q_{hj}^{1*},\eta_{hj}^{1*}\right)}{\partial q_{hj}^1} \geq M \quad \forall Q^2, q_{hj}^1 \geq N, \forall h, j \tag{15}$$

$$\alpha_i \frac{\partial r_{ij}^2\left(q_{ij}^{2*},\eta_{ij}^{2*}\right)}{\partial q_{ij}^2} + \frac{\partial c_{ij}^2\left(q_{ij}^{2*},\eta_{ij}^{2*}\right)}{\partial q_{ij}^2} + \frac{\partial g_{ij}^2\left(q_{ij}^{2*},\eta_{ij}^{2*}\right)}{\partial q_{ij}^2} - \frac{\partial e_{ij}^2\left(q_{ij}^{2*}\right)}{\partial q_{ij}^2} + \alpha_j \frac{\partial \hat{r}_{ij}^2\left(q_{ij}^{2*},\eta_{ij}^{2*}\right)}{\partial q_{ij}^2}$$
$$+ \frac{\partial c_j\left(Q^{3*}\right)}{\partial q_{ij}^2} + \frac{\partial \hat{c}_{ij}^2\left(q_{ij}^{2*},\eta_{ij}^{2*}\right)}{\partial q_{ij}^2} + \frac{\partial \hat{g}_{ij}^2\left(q_{ij}^{2*},\eta_{ij}^{2*}\right)}{\partial q_{ij}^2} \geq M \quad \forall Q^3, q_{ij}^2 \geq N, \forall i, j \tag{16}$$

$$w_{2i}^2\left(r^i(q_i^*)\right) \frac{\partial r^i(q_i^*)}{\partial q_{ik}^2} + \frac{\partial w_{2i}^2\left(r^i(q_i^*)\right)}{\partial q_{ik}^2} r^i(q_i^*) + \frac{\partial c_{ik}^2\left(q_{ik}^{2*}\right)}{\partial q_{ik}^2} + \frac{\partial g_{ik}^2\left(q_{ik}^{2*}\right)}{\partial q_{ik}^2} - \frac{\partial e_{ik}^2\left(q_{ik}^{2*}\right)}{\partial q_{ik}^2}$$
$$+ \hat{c}_{ik}^2\left(Q^{4*},Q^{5*}\right) + \hat{g}_{ik}^2\left(q_{ik}^{2*},\eta_{ik}^{2*}\right) \geq M \quad \forall Q^4, q_{ik}^2 \geq N, \forall i, k \tag{17}$$

$$\alpha_j \frac{\partial r_{jk}^3\left(q_{jk}^{3*},\eta_{jk}^{3*}\right)}{\partial q_{jk}^3} + \frac{\partial c_{jk}^3\left(q_{jk}^{3*},\eta_{jk}^{3*}\right)}{\partial q_{jk}^3} + \frac{\partial g_{jk}^3\left(q_{jk}^{3*},\eta_{jk}^{3*}\right)}{\partial q_{jk}^3} - \frac{\partial e_{jk}^3\left(q_{jk}^{3*}\right)}{\partial q_{jk}^3}$$
$$+ \hat{c}_{jk}^3\left(Q^{4*},Q^{5*},H^{4*},H^{5*}\right) + \hat{g}_{jk}^3\left(q_{jk}^{3*},\eta_{jk}^{3*}\right) \geq M \quad \forall Q^5, q_{jk}^3 \geq N, \forall j, k \tag{18}$$

$$d_k\left(\rho^{4*}\right) \leq M \quad \forall \rho_k^4 > R, \forall k \tag{19}$$

In summary, the unique condition of the variational inequality equilibrium solution is: assuming that the vector function F in equations (32), (33), (34), and (35) are strictly monotonic with respect to $(Q^1, Q^2, Q^3, Q^4, Q^5, H^1, H^2, H^3, H^4, H^5, \rho^4)$, then there must be a unique fund streaming $(Q^{1*}, Q^{2*}, Q^{3*}, Q^{4*}, Q^{5*})$ and unique social network relationship levels $(H^1, H^2, H^3, H^4, H^5)$ and unique demand price vector $\rho^{4*}$ satisfying the equilibrium conditions of the entire financial market.

In order to verify the validity of the above-mentioned financial market super-network model, without loss of generality, we have established a financial super network simulation model with two fund owners, two Internet financial platforms, two traditional financial platforms and three capital demanders. Using MATLAB to implement the projection dynamic system algorithm, the convergence criterion is that the difference between the flow rate and price of each layer in the successive iteration process is less than $\varepsilon = 10^{-4}$. After 3187 iterations, the convergence effect is superior as shown in Figure 7. The good convergence effect verifies the validity of the model.

Consider the optimal equilibrium solution after convergence of the super network model under social network and internet finance as follows:

$Q^{1*}: q_{11}^{1*} = 11.47; q_{12}^{1*} = 11.45; q_{21}^{1*} = 11.47; q_{22}^{1*} = 11.45$

$Q^{2*}: q_{11}^{1*} = 4.44; q_{12}^{1*} = 4.64; q_{21}^{1*} = 4.44; q_{22}^{1*} = 4.64$

$Q^{3*}: q_{11}^{2*} = 11.39; q_{12}^{2*} = 0; q_{21}^{2*} = 0; q_{22}^{2*} = 10.49$

$Q^{4*}: q_{11}^{2*} = 3.31; q_{12}^{2*} = 0; q_{13}^{2*} = 8.24; q_{21}^{2*} = 4.31; q_{22}^{2*} = 8.10; q_{23}^{2*} = 0$

$Q^{5*}: q_{11}^{3*} = 13.50; q_{12}^{3*} = 0; q_{13}^{3*} = 6.77; q_{21}^{3*} = 0; q_{22}^{3*} = 13.41; q_{23}^{3*} = 6.36$.

The clearance price of all funds invested by the Internet financial platform is
$$\gamma_1^* = 294.08; \gamma_2^* = 293.64$$

The clearing price of all the funds of the traditional financial platform when investing is:

$\gamma_1^* = 206.09; \gamma_2^* = 205.65$.

The demand price of the capital demander is:
$\rho_1^{4*} = 471.40; \rho_2^{4*} = 471.30; \rho_3^{4*} = 471.38$.

At this time $S_1^* = S_2^* = 0$, there is no capital flowing to the no investment option, which means that the best decision for the fund owner is to invest all the funds.

It can be seen from $Q^{1*} = 45.84; Q^{2*} = 18.16$ that in the Internet financial environment, fund owners invest more in Internet financial platforms than traditional financial platforms.

For comparative analysis, the adjustment parameters are calculated to obtain the optimal equilibrium solution of the financial super network without considering the social network relationship level as follows:

$Q^{1*}: q_{11}^{1*} = 11.47; q_{12}^{1*} = 11.45; q_{21}^{1*} = 11.47; q_{22}^{1*} = 11.45$

$Q^{2*}: q_{11}^{1*} = 4.44; q_{12}^{1*} = 4.64; q_{21}^{1*} = 4.44; q_{22}^{1*} = 4.64$

$Q^{3*}: q_{11}^{2*} = 11.36; q_{12}^{2*} = 0; q_{21}^{2*} = 0; q_{22}^{2*} = 10.47$

$Q^{4*}: q_{11}^{2*} = 8.24; q_{12}^{2*} = 3.34; q_{13}^{2*} = 0; q_{21}^{2*} = 0; q_{22}^{2*} = 4.32; q_{23}^{2*} = 8.11$

$Q^{5*}: q_{11}^{3*} = 0; q_{12}^{3*} = 6.76; q_{13}^{3*} = 13.48; q_{21}^{3*} = 13.40; q_{22}^{3*} = 6.35; q_{23}^{3*} = 0$.

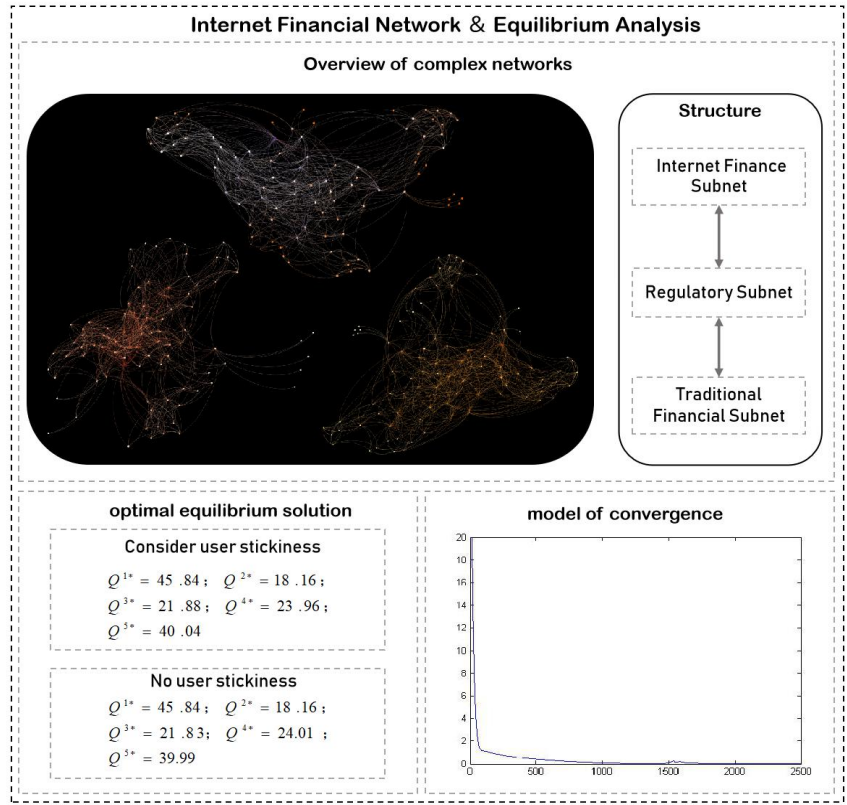

**Fig.** The comparison shows that in both cases, $Q^{1*}$ and $Q^{2*}$ have no change, and the level of social network relationship has no impact on the overall investment decision of the fund owner; $Q^{3*}$, $Q^{4*}$ and $Q^{5*}$ have changes, the level of social network relationship to the Internet financial platform and traditional financial platform investment Decision making has a certain impact. The direct investment of the Internet financial platform in the capital demand market has declined, and the indirect investment through the traditional financial platform has increased, indicating that the traditional financial platform, the Internet financial platform, and the social network relationship of the capital demanders will be strengthened, and the financial market will be more effective.

## 8. Conclusion and suggestion

The results of Internet Finance subnet show that the main source of systemic risk lies in the external effect of microeconomic risk-laking activity. That is, the losses imposed on the whole society by a single firm (institution) are far greater than the losses suffered by the investors themselves. In practical terms, companies with strong power in Internet Finance platforms tend to pay close attention to their own risk prevention and control. Instead, ordinary Internet Finance platforms that do business directly with these platforms are not paying enough attention to their own risks. These sections that lie outside of important nodes. Weak enough and not independent enough, regulation without attention is often the source of risks to Internet Finance. Therefore, for any node in the Internet Finance system, the risk is not to select the node with the greatest degree of connectivity, but to establish a connection with the node with the largest degree of connectivity. This is enough to reflect the diffusivity of risk in Internet Finance risk system and the difference between other complex systems.

In addition, the Internet Finance system has a core impact. Internet giants monopolize more user traffic in the market, which makes the Internet industry with overall long tail risk. Judging from the investor nature of Internet Finance products, most of the investors involved in Internet Finance products lack the awareness of risk prevention. One of the reasons is that the Internet Finance platform promotes the incentive mechanism on a large scale, the other is that most of the Internet Finance products ignore the red tape authentication of traditional investment behavior, thus increasing the core risk of Internet Finance. Therefore, in view of the supervision of Internet Finance, we need to focus on the large-scale Internet gold. Financial platform business activities. The most hierarchical feature of the Internet Finance system is the regulatory layer, which has the key to the overall boom or decline of the network. Therefore, the implementation of supervision is an important factor to control the occurrence of systemic risks in Internet Finance.

Finally, the Internet Finance system has multiple levels of relevance. The systemic risk of Internet Finance has obvious multi-level nature, which is mainly composed of three layers: supervision layer, Internet Finance inner layer and traditional financial layer. The superposition of multi-source risk is originated from the inner layer of Internet Finance and infects or induces the systemic risk of traditional financial industry. In the supervision of the systemic characteristics of Internet Finance, we need to keep any part of the system can not be separated from the whole, targeted control. Internet Finance systems integrate and change each other to cope with external shocks and endogenous fluctuations. Thus, the whole system is approximately in a more stable state. Such self-regulation includes small changes, such as the regulatory level through the continuous optimization of the system to make the entire network to reach a stable state not to be attacked; However, the self-regulation may also come from a larger vibration, such as the damage of a core node, which results in the loss of most of the network structure, so that the network will eventually have to stay in a stable state that is much smaller than before. Often the latter regulation is accompanied by the loss of a large area of nodes, the overall system is weak, the links between nodes have broken, that is to say, the systematic risk of Internet Finance has occurred.

This paper synthesizes the research literature of different scholars, selects 90 representative risk factors of Internet Finance, and puts forward the risk contagion network model of Internet Finance with risk factors as nodes. Considering the Internet Finance is a complex system, for some hard to quantify the risk factor of the problem in the processing, the model has a more intuitive visual ability, help to combine micro and macro two angles to analyze risk contagion of systemic risk, provide a better basis for the implementation of the regulation means. In addition, the study of Internet Finance with the theory of complexity science can become a new idea, new method and new approach, so as to achieve the unification of Internet Finance research on micro and macro level, and provide more powerful support for the supervision of modern Internet Finance and the detection of potential hidden dangers. The subsequent research can collect the transaction data of commercial Banks and Internet Finance enterprises for empirical analysis, so as to explore the contagion law of systemic risk.

## Acknowledgments

We thank the anonymous reviewers for their careful reading of our manuscript and their many insightful comments and suggestions. This work is supported by the National Social Science Foundation of China (17BGL055). The first author also thanks his great father, who gave him a lot of encouragements at the end of his life.

## References

[1] Adrian, T., & Brunnermeier, M. (2011). CoVaR. doi:10.3386/w17454
[2] Huang, X., Zhou, H., & Zhu, H. (2012). Assessing the systemic risk of a heterogeneous portfolio of banks during the recent financial crisis. Journal of Financial Stability, 8(3), 193–205. doi:10.1016/j.jfs.2011.10.004
[3] Acharya, V., Engle, R., & Richardson, M. (2012). Capital Shortfall: A New Approach to Ranking and Regulating Systemic Risks. American Economic Review, 102(3), 59–64. doi:10.1257/aer.102.3.59
[4] Jobst, A. A., & Gray, D. F. (2013). Systemic Contingent Claims Analysis: Estimating Market-Implied Systemic Risk. IMF Working Papers, 13(54), 1. doi:10.5089/9781475572780.001
[5] Lee, E., & Lee, B. (2012). Herding behavior in online P2P lending: An empirical investigation. Electronic Commerce Research and Applications, 11(5), 495–503. doi:10.1016/j.elerap.2012.02.001
[6] Kotha, S., Rajgopal, S., & Rindova, V. (2001). Reputation Building and Performance: An Empirical Analysis of the Top-50 Pure Internet Firms. European Management Journal, 19(6), 571–586. doi:10.1016/s0263-2373(01)00083-4
[7] Cont, R., Moussa, A., & Santos, E. B. e. (2010). Network Structure and Systemic Risk in Banking Systems. SSRN Electronic Journal. doi:10.2139/ssrn.1733528
[8] Mistrulli, P. E. (2011). Assessing financial contagion in the interbank market: Maximum entropy versus observed interbank lending patterns. Journal of Banking & Finance, 35(5), 1114–1127. doi:10.1016/j.jbankfin.2010.09.018
[9] Jalili, M., & Perc, M. (2017). Information cascades in complex networks. Journal of Complex Networks. doi:10.1093/comnet/cnx019
[10] Kenett, D. Y., Perc, M., & Boccaletti, S. (2015). Networks of networks – An introduction. Chaos, Solitons & Fractals, 80, 1–6. doi:10.1016/j.chaos.2015.03.016
[11] Georg, C.-P. (2013). The effect of the interbank network structure on contagion and common shocks. Journal of Banking & Finance, 37(7), 2216–2228. doi:10.1016/j.jbankfin.2013.02.032
[12] João Barata Ribeiro Blanco Barroso, Thiago Christiano Silva, Sergio Rubens Stancato de Souza. (2018). Identifying systemic risk drivers in financial networks, Physica A: Statistical Mechanics and its Applications, 503: 650-674. doi: 10.1016 / j.physa.2018.02.144
[13] Solorzano-Margain, J. P., Martinez-Jaramillo, S., & Lopez-Gallo, F. (2013). Financial contagion: extending the exposures network of the Mexican financial system. Computational Management Science, 10(2-3), 125–155. doi:10.1007/s10287-013-0167-5
[14] Amini, H., & Minca, A. (2016). Inhomogeneous Financial Networks and Contagious Links. Operations Research, 64(5), 1109–1120. doi:10.1287/opre.2016.1540
[15] Smolyak, A., Levy, O., Shekhtman, L., & Havlin, S. (2018). Interdependent networks in Economics and Finance—A Physics approach. Physica A: Statistical Mechanics and Its Applications, 512, 612–619. doi:10.1016/j.physa.2018.08.089
[16] Watts, D. J., & Strogatz, S. H. (1998). Collective dynamics of "small-world" networks. Nature, 393(6684), 440–442. doi:10.1038/30918
[17] Podobnik, B., Horvatic, D., Lipic, T., Perc, M., Buldú, J. M., & Stanley, H. E. (2015). The cost of attack in competing networks. Journal of The Royal Society Interface, 12(112), 20150770. doi:10.1098/rsif.2015.0770
[18] Leduc, M. V., & Thurner, S. (2017). Incentivizing Resilience in Financial Networks. SSRN Electronic Journal . doi:10.2139/ssrn.2794371
[19] Battiston, S., Puliga, M., Kaushik, R., Tasca, P., & Caldarelli, G. (2012). DebtRank: Too Central to Fail? Financial Networks, the FED and Systemic Risk. Scientific Reports, 2(1). doi:10.1038/srep00541
[20] Schweitzer, F., Fagiolo, G., Sornette, D., Vega-Redondo, F., Vespignani, A., & White, D. R. (2009). Economic Networks: The New Challenges. Science, 325(5939), 422–425. doi:10.1126/science.1173644
[21] Hidalgo, C. A., Klinger, B., Barabasi, A.-L., & Hausmann, R. (2007). The Product Space Conditions the Development of Nations. Science, 317(5837), 482–487. doi:10.1126/science.1144581
[22] Leduc, M., Poledna, S., & Thurner, S. (2017). Systemic risk management in financial networks with credit default swaps. The Journal of Network Theory in Finance. doi:10.21314/jntf.2017.034
[23] Sakamoto, Y., & Vodenska, I. (2017). Erratum to "Systemic risk and structural changes in a bipartite bank network: a new perspective on the Japanese banking crisis of the 1990s." Journal of Complex Networks, 5(3), 512–512. doi:10.1093/comnet/cnx012
[24] Ronghua Xu, Wing-Keung Wong, Guanrong Chen & Shuo Huang: Topological Characteristics of the Hong Kong Stock Market: A Test-based P-threshold Approach to Understanding Network Complexity. Sci. Rep.7, 41379 (2017).
[25] Bonanno, G., Caldarelli, G., Lillo, F., Miccich, S., Vandewalle, N., & Mantegna, R. N. (2004). Networks of equities in financial markets. The European Physical Journal B - Condensed Matter, 38(2), 363–371. doi:10.1140/epjb/e2004-00129-6
[26] Watts, D. J., & Strogatz, S. H. (1998). Collective dynamics of "small-world" networks. Nature, 393(6684), 440–442. doi:10.1038/30918
[27] Barabási, A. (1999). Emergence of Scaling in Random Networks. Science, 286(5439), 509–512. doi:10.1126/science.286.5439.509
[28] Newman, M. E. J., Strogatz, S. H., & Watts, D. J. (2001). Random graphs with arbitrary degree distributions and their applications. Physical Review E, 64(2). doi:10.1103/physreve.64.026118
[29] Strogatz, S. H. (2001). Exploring complex networks. Nature, 410(6825), 268–276. doi:10.1038/35065725
[30] Poledna, S., Molina-Borboa, J. L., Martínez-Jaramillo, S., van der Leij, M., & Thurner, S. (2015). The multi-layer network nature of systemic risk and its implications for the costs of financial crises. Journal of Financial Stability, 20, 70–81. doi:10.1016/j.jfs.2015.08.001
[31] Newman, M. E. J. (2004). Detecting community structure in networks. The European Physical Journal B - Condensed Matter, 38(2), 321–330. doi:10.1140/epjb/e2004-00124-y
[32] Dey, P., & Roy, S. (2015). A Comparative Analysis of Different Social Network Parameters Derived from Facebook Profiles. Proceedings of the Second International Conference on Computer and Communication Technologies, 125–132. doi:10.1007/978-81-322-2517-1_13
[33] Jia nan. Research on the impact of Internet Finance on banking risks in China and its systematic risk measurement [J]. Exploration of economic problems, 2018, (04):145-157. (in Chinese)
[34] Xiaohui Hou, Zhixian Gao, Qing Wang, Internet Finance development and banking market discipline: Evidence from China, Journal of Financial Stability, Volume 22, 2016, Pages 88-100, ISSN 1572-3089. (in Chinese)
[35] Yin Haiyuan,Wang Panpan. Current situation and system construction of Internet Finance supervision in China [J]. Science of finance and economics,2015(09):12-24. (in Chinese)
[36] Zhu Chen, Hua Guihong. The impact of Internet Finance on the systemic risk of China's banking industry -- an empirical study based on SCCA model and stepwise regression method [J]. Financial economics research,2018,33(02):50-59. (in Chinese)
[37] Liu Zhonglu, Lin Zhangyue.Research on the impact of Internet Finance on the profitability of commercial Banks [J]. Beijing social science, 2016(9). (in Chinese)

[38] Debreceny, R., Gray, G. L., & Rahman, A. (2002). The determinants of Internet Finance reporting. Journal of Accounting and Public Policy, 21(4-5), 371–394. doi:10.1016/s0278-4254(02)00067-4

[39] Smedlund, A. (2012). Value Cocreation in Service Platform Business Models. Service Science, 4(1), 79–88. doi:10.1287/serv.1110.0001

[40] Social networking relationships, firm-specific managerial experience and firm performance in a transition economy: A comparative analysis of family owned and nonfamily firms

[41] Laeven, L. A., & Levine, R. (2008). Bank Governance, Regulation, and Risk Taking. SSRN Electronic Journal. doi:10.2139/ssrn.1142967

[42] Chaffee, E. C., & Rapp, G. C. (2012). Regulating Online Peer-to-Peer Lending in the Aftermath of Dodd-Frank: In search of an evolving regulatory regime for an evolving industry. Wash. & Lee L. Rev., 69, 485.http://scholarlycommons.law.wlu.edu/wlulr/vol69/iss2/4

[43] Haldane, A. G., & May, R. M. (2011). Systemic risk in banking ecosystems. Nature, 469(7330), 351–355. doi:10.1038/nature09659

[44] Glasserman, P., & Young, H. P. (2015). How likely is contagion in financial networks? Journal of Banking & Finance, 50, 383–399. doi:10.1016/j.jbankfin.2014.02.006

[45] Zuo, Z., & Chen, M. (2011). Analysis of the Behavior of Fund Investment in China on the Basis of LSV Herding Measure. 2011 Fourth International Conference on Business Intelligence and Financial Engineering. doi:10.1109/bife.2011.18

[46] Weber, G. W., Kropat, E., Tezel, A., & Belen, S. (2010). Optimization applied on regulatory and eco-finance networks-survey and new developments. Pac. J. Optim, 6(2), 319-340.

[47] Kropat, E., Weber, G. W., & Akteke-Öztürk, B. (2008). Eco-finance networks under uncertainty. In Proceedings of the international conference on engineering optimization.

[48] Kara, G., Özmen, A., & Weber, G. W. (2019). Stability advances in robust portfolio optimization under parallelepiped uncertainty. Central European Journal of Operations Research, 27(1), 241-261.

[49] Bekaert, G., Hoerova, M., & Duca, M. L. (2013). Risk, uncertainty and monetary policy. Journal of Monetary Economics, 60(7), 771-788.

[50] Cruz, J. M., & Wakolbinger, T. (2008). Multiperiod effects of corporate social responsibility on supply chain networks, transaction costs, emissions, and risk. International journal of production economics, 116(1), 61-74.

[51] Nagurney, A., Cruz, J., & Wakolbinger, T. (2007). The co-evolution and emergence of integrated international financial networks and social networks: Theory, analysis, and computations. In Globalization and Regional Economic Modeling (pp. 183-226). Springer, Berlin, Heidelberg.

[52] Gai, P., & Kapadia, S. (2010). Contagion in financial networks. Proceedings of the Royal Society A: Mathematical, Physical and Engineering Sciences, 466(2120), 2401-2423.

[53] Bougheas, S., & Kirman, A. (2015). Complex financial networks and systemic risk: A review. In Complexity and Geographical Economics (pp. 115-139). Springer, Cham.

[54] Aldasoro, I., Gatti, D. D., & Faia, E. (2017). Bank networks: Contagion, systemic risk and prudential policy. Journal of Economic Behavior & Organization, 142, 164-188.